\documentclass[a4paper,12pt]{article}
\usepackage{epsfig}
\usepackage{multirow}

\def\beq{\begin{equation}}
\def\eeq{\end{equation}}
\def\bea{\begin{eqnarray}}
\def\eea{\end{eqnarray}}
\def\psl{\hbox{\hbox{${p}$}}\kern-1.9mm{\hbox{${/}$}}}

\def\dofigB#1#2{\centerline{
\epsfxsize=#1\epsfig{file=#2, width=8cm,height=8cm, angle=-90}
\hspace{0cm}
}}

\def\dofigA#1#2{\centerline{\epsfxsize=#1\epsfig{file=#2, width=15cm, 
height=10cm, angle=0}
\vspace{-1cm}
}}

\def\dofigs#1#2#3{\centerline{\epsfxsize=#1\epsfig{file=#2, width=8cm, 
height=8cm, angle=-90}
\hfil\epsfxsize=#1\epsfig{file=#3,  width=8cm, height=8cm, angle=-90}}}

\newcommand{\text}{\rm}

\newcommand{\YFLR}{{\bf Y^{L,R}_{F}}}
\newcommand{\YDL}{{\bf Y^{L}_{D}}}
\newcommand{\YDR}{{\bf Y^{R}_{D}}}
\newcommand{\YUL}{{\bf Y^{L}_{U}}}
\newcommand{\YUR}{{\bf Y^{R}_{U}}}
\newcommand{\SM}{{\rm \scriptscriptstyle SM}}
\newcommand{\YSM}{Y^{\SM}}

\def\lsim{\raise0.3ex\hbox{$\;<$\kern-0.75em\raise-1.1ex\hbox{$\sim\;$}}} 

\def\gsim{\raise0.3ex\hbox{$\;>$\kern-0.75em\raise-1.1ex\hbox{$\sim\;$}}}

\begin{document}
\begin{titlepage}

\vskip 2cm
\begin{center} 

\boldmath
{\Large\bf 
Effective Yukawa couplings and flavor-changing

\vspace{0.2cm}
Higgs boson decays  at linear colliders}

\vspace{1cm}

E. Gabrielli$^a$ and B. Mele$^b$
\vskip 1.cm

\emph{
$^a$ CERN, PH-TH, CH-1211 Geneva 23, Switzerland\\
$^b$ INFN, Sezione di Roma, c/o Dip. di Fisica,
 Universit\`a di Roma  ``La Sapienza'',\\ Piazzale A. Moro 2, I-00185 Rome, Italy
}
\end{center}

\vskip 0.7cm
\begin{abstract}
We analyze 
the advantages of a linear-collider program for testing 
a recent theoretical proposal
where the Higgs-boson Yukawa couplings are radiatively generated, keeping unchanged the standard-model mechanism for   electroweak-gauge-symmetry breaking. Fermion masses arise  at a large energy scale through an unknown mechanism, and 
the standard model at the electroweak scale is  regarded as an effective field theory.
In this scenario,  Higgs boson  
decays into photons and electroweak gauge-boson pairs are considerably enhanced for a light Higgs boson, which makes
a signal observation at the LHC straightforward. On the other hand,
the clean environment of a linear collider is required to directly probe the radiative fermionic sector of the Higgs boson 
couplings.
Also, we show that the flavor-changing 
Higgs boson decays are  dramatically enhanced with respect to the standard model. In particular,  we find  a measurable  
branching ratio in the range $(10^{-4}-10^{-3})$  for the decay $H\to bs$  for a Higgs boson lighter than 140 GeV, depending on the high-energy scale where Yukawa couplings vanish. 
We present a detailed analysis of the Higgs boson production cross sections 
 at  linear colliders for interesting decay signatures, as well as branching-ratio correlations for different 
flavor-conserving/nonconserving fermionic decays.

\end{abstract}
\end{titlepage}

\section{Introduction}

The clarification of the electroweak symmetry breaking (EWSB) mechanism 
is the most urgent
task of the Large Hadron Collider (LHC), that last year started taking data at an unprecedented $pp$ collision energy of $\sqrt{S}=7$ TeV. With more collected integrated luminosity and a possible collision energy upgrade, this might soon lead to the 
long-awaited discovery of the Higgs boson  \cite{Djouadi}.
While the observation and study of the properties of a scalar particle with features not too different from the ones of the 
standard-model (SM) Higgs boson  will be accessible at the LHC, it is well known that a detailed study of the Higgs-boson profile and couplings will crucially benefit from a future  $e^+e^-$ 
linear-collider program \cite{LC,clic}.

In \cite{our}, we introduced a new phenomenological framework giving  an {\it improved} description of the fermiophobic (FP) Higgs-boson scenario \cite{HiggsFP}. In particular,
we considered the possibility that the Higgs boson gives mass to the electroweak (EW) vector bosons just as in the SM, while fermion masses and chiral-symmetry breaking (ChSB) arise from a different unknown mechanism at an energy scale considerably larger than  $M_W$. Then, the Higgs boson is coupled to the EW vector bosons just as in the SM, while 
Higgs Yukawa couplings are missing at tree-level in the fermion Lagrangian. 
Yukawa couplings are anyway generated at one loop after ChSB is introduced by {\it nonstandard} explicit fermion mass terms in the Lagrangian. One new energy parameter 
 $\Lambda\sim 10^{(4-16)}$ GeV (the renormalization scale where the renormalized Yukawa couplings vanish) is introduced to give an effective description of the radiative effects of ChSB on Higgs couplings to fermions at low energies. 
Important logarithmic effects for large values of $\Lambda$  are  resummed  via renormalization-group (RG) equations in \cite{our}.

Radiative Higgs couplings to fermions turn out in general to be smaller than the corresponding tree-level SM Yukawa couplings. For instance, for $m_H<160$ GeV, the {\it effective} Yukawa coupling to $b$ quarks is about 20 to 5 times smaller than the corresponding SM value for $\Lambda= 10^{4}$ GeV to $10^{16}$ GeV.
Nevertheless, the simultaneous reduction in the Higgs boson width, corresponding to the depleted  coupling to fermions, considerably compensates for the decrease of the fermionic Higgs decay widths,
and gives quite enhanced  radiative Higgs branching ratios (BR's) to fermions.
For $\Lambda\sim 10^{16}$ GeV and  $m_H\lsim 130$ GeV,  one gets  branching ratios to the $b$ quarks comparable to the SM values.

In \cite{our}, we also discussed the phenomenological expectations at the LHC for the present theoretical framework. Because of the suppression of the Higgs-gluon  effective coupling $ggH$ following the absence of the tree-level top-quark Yukawa coupling, the  Higgs-boson production at the LHC occurs predominantly by vector-boson fusion (VBF) and associated $WH/ZH$ production (VH) with SM cross sections. For $m_H<150$ GeV, the decay BR's for the channels  $H\to \gamma\gamma,WW,ZZ,Z\gamma$ can be  enhanced  with respect to their SM values by as much as an order of magnitude or more, because of the depleted Higgs total width \cite{our}. As a consequence, in the present scenario, an enhanced two-photon resonance signal in the VBF and $WH/ZH$ production could easily emerge from the background. Indeed, the additional jets (or leptons) in the final states would crucially help in pinpointing  the signal events with respect to the SM case, where the dominant  production is through $gg\to H$. The study of the decay channels 
$H\to \gamma\gamma,WW,ZZ$ at the LHC will give enough information to start to shape up the effective Yukawa scenario  with some sensitivity to the scale $\Lambda$. On the other hand, the study of the complementary fermion decay channels $H\to f\bar f$, that are very challenging at the LHC even in the easier SM case, will require the clean environment of a linear-collider program.

In  this paper, we   discuss  how the excellent potential of a linear collider machine for the  precision measurements of the Higgs couplings to fermions could help in testing their radiative structure as predicted in the  effective-Yukawa framework. For the  first time, beyond the flavor-diagonal $H f\bar f$ couplings,  we will go through the flavor-changing (FC) sector of the model, where we find large enhancements with respect to the SM predictions. We will show that studies of the FC Higgs-boson decays at linear colliders can provide extra handles to consolidate the effective Yukawa framework.

The plan of the paper is the following. In Sec. 2,  the basic phenomenological features of the effective-Yukawa model are reviewed. In Sec. 3, Higgs-boson production cross sections in $e^+e^-$ collisions at the c.m. energy 
$\sqrt{S}=350$ GeV are presented for different Higgs decay channels. Correlations between the BR's for the most important  fermionic Higgs decays are shown.
In Sec. 4, FC Higgs couplings are computed via 
RG equations. FC decay BR's are discussed in Sec.5.
 Our conclusions are presented in Sec. 6.

\section{The effective-Yukawa model}

In this section, we sum up 
the main phenomenological features of the effective-Yukawa model,
as introduced in \cite{our}\footnote{Throughout the paper, for all the basic physical constants and parameters, we assume the same numerical values as in \cite{our}.}. 

In the effective-Yukawa model, 
EW vector bosons acquire mass via spontaneous symmetry breaking just as in the SM, and a physical Higgs boson is left in the spectrum which is coupled to vector bosons via SM couplings.
The peculiar feature of the model is that fermion masses are not assumed to arise from the EW symmetry-breaking mechanism, but from an 
unknown mechanism at an energy scale considerably larger than  $M_W$.  As a consequence,  Higgs Yukawa couplings are missing at tree level  in the fermion Lagrangian.
They are anyway radiatively  generated at one-loop after 
ChSB is introduced by {\it non standard} explicit fermion mass terms in the Lagrangian.

In the model, there is just one new free parameter, the energy scale  $\Lambda$, defined  as the renormalization scale 
where all the Yukawa matrix elements (in flavor space) are assumed to vanish. 
This renormalization condition just sets 
the Higgs-fermion decoupling at the high-energy  $\Lambda\gg M_W$. In particular, we consider 
$\Lambda$ in the range $10^{(4-16)}$ GeV.
Large logarithmic  contributions 
$g_i^{2n }\log^n{(\Lambda/m_H)}$
(where $g_i$ are the SM gauge couplings) 
to the Yukawa operators are then expected at higher orders 
in perturbation theory that 
  can be resummed via the standard technique of the 
RG equations.
Notice that the coefficients multiplying these log-terms are universal, that is independent of the structure of the UV completion of the theory. 
Therefore, they can be calculated in the corresponding effective theory 
by evaluating the anomalous-dimension matrix of the Yukawa couplings.

\begin{figure}[tpb]
\begin{center}
\dofigB{3.1in}{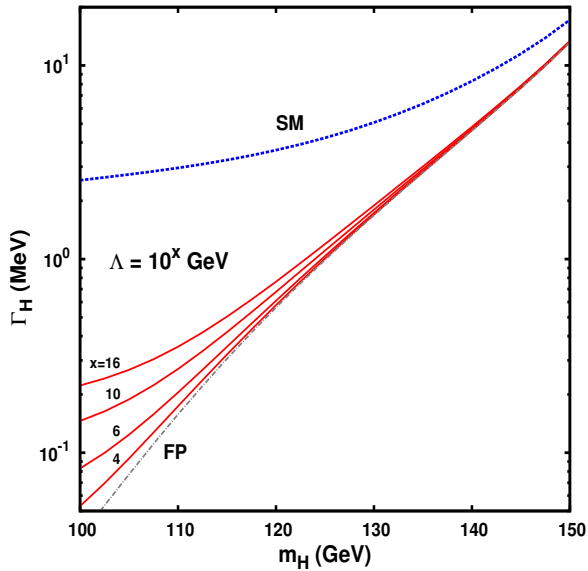}
\end{center}
\caption{\small 
Total Higgs-boson width  versus $m_H$, for different values of $\Lambda$. The curves labeled by SM and FB correspond to the standard-model and the naive fermiophobic Higgs scenarios, respectively.
 }
\label{fig1}
\end{figure}

As anticipated, while 
radiative Higgs couplings to fermions in this scenario are smaller than the corresponding SM Yukawa couplings, BR's for fermionic Higgs decays can be conspicuous for large $\Lambda$ and 
$m_H\lsim 140$ GeV. 
Indeed, the suppression of the fermionic Higgs couplings and of the related fermionic Higgs widths is compensated for by the corresponding depletion in the  total Higgs-boson width. 
In Fig.\ref{fig1}, the total Higgs-boson width is shown versus $m_H$ for different values of $\Lambda$. 

\noindent
In Fig.\ref{fig1}, and in all subsequent plots and tables,
the SM and FP labels stand for 
the  standard-model  and  the naive fermiophobic Higgs scenario results, respectively\footnote{We define as {\it naive fermiophobic Higgs} scenario, a model where all the Higgs fermionic couplings are assumed to vanish at the EW scale, and the Higgs boson is coupled to vector bosons as in the SM.}.
Because of the  fall in the light-Higgs total width, values of  
BR$(H\to b\bar b)$ as large as  the SM ones can be obtained at high $\Lambda$'s (cf.  Table~\ref{table1}, taken from \cite{our}). 
In particular,
for $m_H\simeq (100-130)$ GeV  and $\Lambda\simeq 10^{16}$ GeV,  one gets    BR$(H\to b\bar b)\simeq (82-10)\%$ from radiative effects, to be compared with the corresponding SM values BR$(H\to b\bar b)\simeq(82-54)\%$.  
\begin{table}[htbp]
\begin{center}
\begin{tabular}{|c|c|c|c|c|c|c|c|c|}                
 \hline {\large
 $m_H$} & {$\Lambda$} &  $\gamma\gamma$ & $WW$ &  $ZZ$ & $Z\gamma$ & $b\bar{b}$ & $c\bar{c}$ & $\tau\bar{\tau}$ \\ \hline
 $({\rm GeV})$ & $({\rm GeV})$ & BR(\%) & BR(\%)  & BR(\%) & BR(\%) & BR(\%) & BR(\%) & BR(\%) \\ \hline \hline                       \multirow{4}{*}{100}                                              & $10^4$ &
  12                                   &
  52                                   &
         5.1                                   &
        0.26                                   &
  30                                   &
        0.15                                   &
       0.076                                  \\
     & $10^6$ &
         8.0                                   &
  33                                   &
         3.3                                   &
        0.17                                   &
  55                                   &
        0.28                                   &
        0.17                                  \\
  & $10^{10}$ &
         4.6                                   &
  19                                   &
         1.9                                   &
       0.094                                   &
  74                                   &
        0.38                                   &
        0.30                                  \\
  & $10^{16}$ &
         3.0                                   &
  12                                   &
         1.2                                   &
       0.062                                   &
  82                                   &
        0.43                                   &
        0.44          \\ \hline \multirow{2}{*}{  100     }
         & FP &
  18                                   &
  74                                   &
         7.4                                   &
        0.37                                   &
         0          &
         0          &
         0         \\
         & SM &
        0.15                                   &
         1.1                                   &
        0.11                                   &
       0.005                                   &
  82                                   &
         3.8                                   &
         8.3                                  \\
                                                                                                                                                                                                                                                                                                                                                                                                                                                                                                                                                                          \hline \hline \multirow{4}{*}{110}  & $10^4$ &
         5.3                                   &
  78                                   &
         7.0                                   &
        0.72                                   &
         9.1                                   &
       0.071                                   &
       0.036                                  \\
     & $10^6$ &
         4.6                                   &
  66                                   &
         5.9                                   &
        0.61                                   &
  22                                   &
        0.18                                   &
        0.11                                  \\
  & $10^{10}$ &
         3.5                                   &
  50                                   &
         4.5                                   &
        0.46                                   &
  41                                   &
        0.33                                   &
        0.26                                  \\
  & $10^{16}$ &
         2.7                                   &
  38                                   &
         3.4                                   &
        0.36                                   &
  54                                   &
        0.45                                   &
        0.45          \\ \hline \multirow{2}{*}{  110     }
         & FP &
         5.8                                   &
  86                                   &
         7.7                                   &
        0.79                                   &
         0          &
         0          &
         0         \\
         & SM &
        0.18                                   &
         4.6                                   &
        0.41                                   &
       0.037                                   &
  78                                   &
         3.6                                   &
         7.9                                  \\
                                                                                                                                                                                                                                                                                                                                                                                                                                                                                                                                                                          \hline \hline \multirow{4}{*}{120}  & $10^4$ &
         2.2                                   &
  85                                   &
         9.4                                   &
        0.75                                   &
         2.6                                   &
       0.032                                   &
       0.016                                  \\
     & $10^6$ &
         2.1                                   &
  81                                   &
         8.9                                   &
        0.72                                   &
         7.5                                   &
       0.092                                   &
       0.056                                  \\
  & $10^{10}$ &
         1.9                                   &
  72                                   &
         8.0                                   &
        0.64                                   &
  17                                   &
        0.21                                   &
        0.16                                  \\
  & $10^{16}$ &
         1.7                                   &
  64                                   &
         7.1                                   &
        0.57                                   &
  26                                   &
        0.32                                   &
        0.33          \\ \hline \multirow{2}{*}{  120     }
         & FP &
         2.3                                   &
  87                                   &
         9.7                                   &
        0.77                                   &
         0          &
         0          &
         0         \\
         & SM &
        0.21                                   &
  13                                   &
         1.5                                   &
        0.11                                   &
  69                                   &
         3.2                                   &
         7.0                                  \\
                                                                                                                                                                                                                                                                                                                                                                                                                                                                                                                                                                          \hline \hline \multirow{4}{*}{130}  & $10^4$ &
         1.0                                   &
  86                                   &
  11                                   &
        0.63                                   &
        0.84                                   &
       0.016                                   &
       0.008                                  \\
     & $10^6$ &
         1.0                                   &
  85                                   &
  11                                   &
        0.62                                   &
         2.6                                   &
       0.048                                   &
       0.029                                  \\
  & $10^{10}$ &
        1.0                                   &
  81                                   &
  11                                   &
        0.59                                   &
         6.1                                   &
        0.12                                   &
       0.092                                  \\
  & $10^{16}$ &
        0.96                                   &
  77                                   &
  10                                   &
        0.57                                   &
  10                                   &
        0.20                                   &
        0.20          \\ \hline \multirow{2}{*}{  130     }
         & FP &
         1.0                                   &
  87                                   &
  11                                   &
        0.63                                   &
         0          &
         0          &
         0         \\
         & SM &
        0.21                                   &
  29                                   &
         3.8                                   &
        0.19                                   &
  54                                   &
         2.5                                   &
         5.4                                  \\
                                                                                                                                                                                                                                                                                                                                                                                                                                                                                                                                                                          \hline \hline \multirow{4}{*}{140}  & $10^4$ &
        0.53                                   &
  87                                   &
  12                                   &
        0.48                                   &
        0.29                                   &
       0.008                                   &
       0.004                                  \\
     & $10^6$ &
        0.53                                   &
  86                                   &
  12                                   &
        0.48                                   &
        0.90                                   &
       0.026                                   &
       0.016                                  \\
  & $10^{10}$ &
        0.53                                   &
  85                                   &
  12                                   &
        0.47                                   &
         2.3                                   &
       0.064                                   &
       0.051                                  \\
  & $10^{16}$ &
        0.52                                   &
  83                                   &
  11                                   &
        0.46                                   &
         4.1                                   &
        0.11                                   &
        0.12          \\ \hline \multirow{2}{*}{  140     }
         & FP &
        0.53                                   &
  87                                   &
  12                                   &
        0.48                                   &
         0          &
         0          &
         0         \\
         & SM &
        0.19                                   &
  48                                   &
         6.6                                   &
        0.24                                   &
  36                                   &
         1.6                                   &
         3.6             \\ \hline \end{tabular}
\end{center} 
\caption[]{Branching ratios (in percentage) for  dominant 
Higgs-boson decays, for different values of 
the Higgs mass and  $\Lambda$ (taken from \cite{our}). The SM and FP labels stand for 
the  standard-model  and  the naive fermiophobic Higgs scenarios, respectively.}
\label{table1} 
\end{table}

\begin{figure}[tpb]
\begin{center}
\dofigs{3.1in}{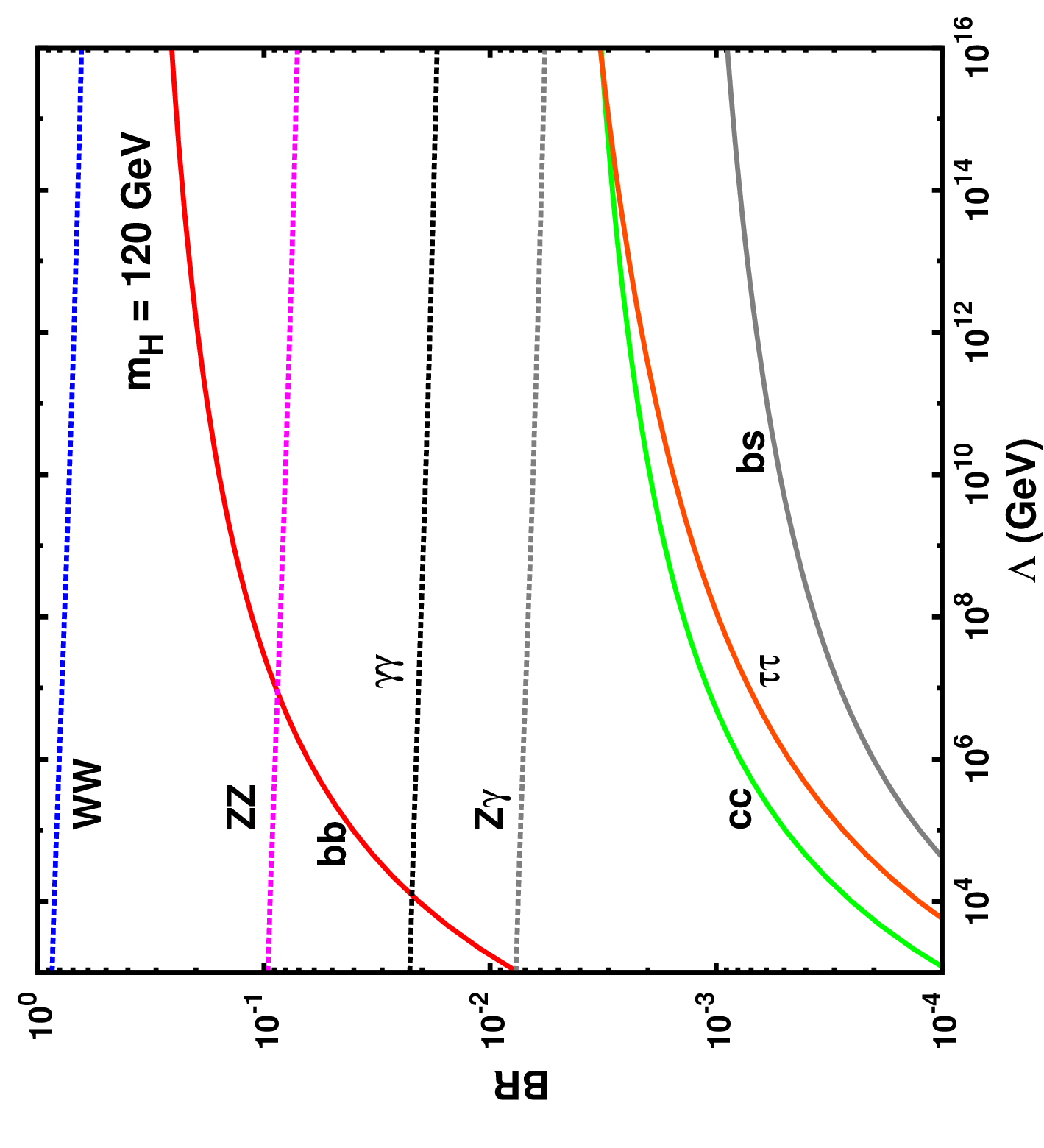}{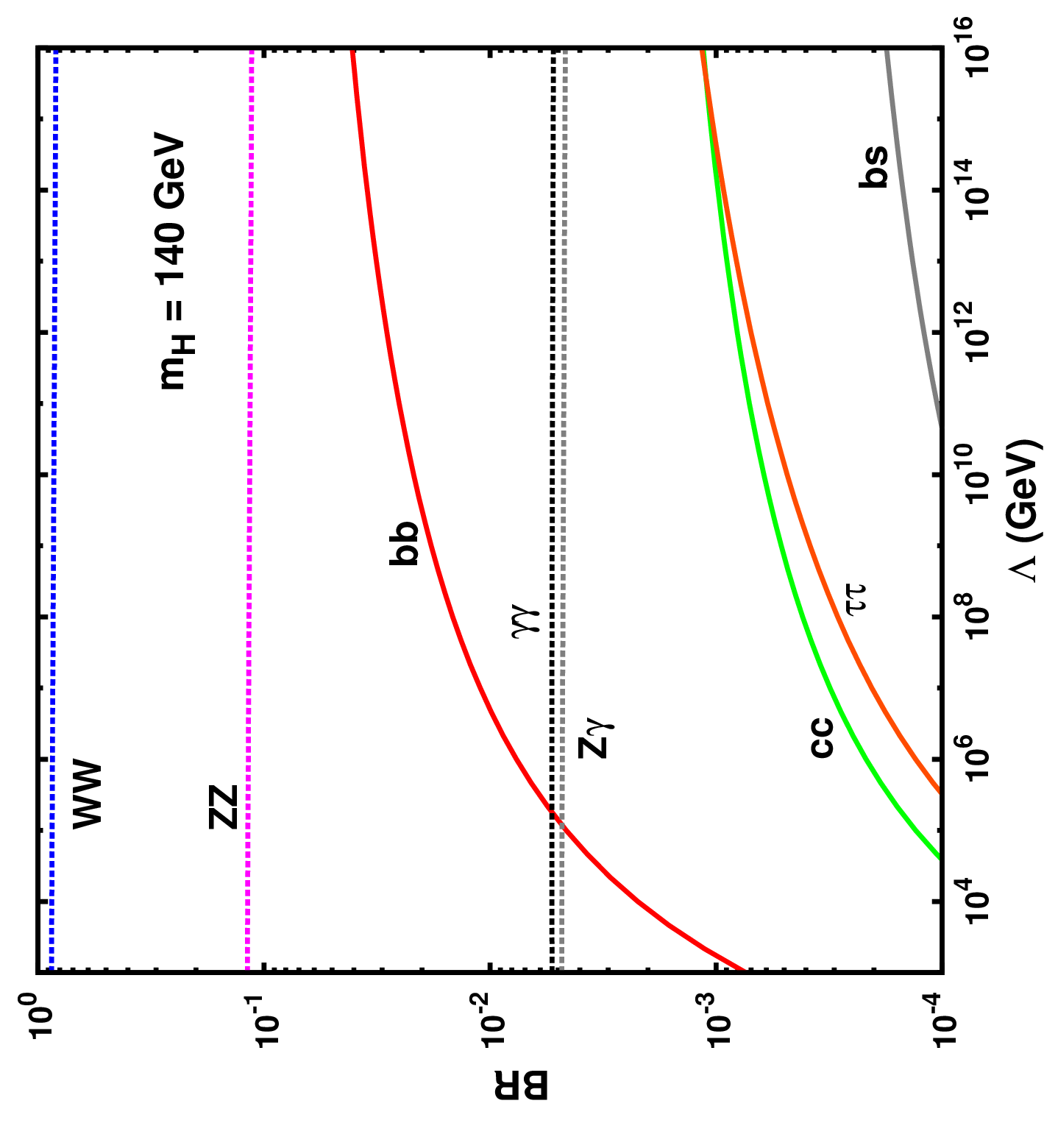}
\end{center}
\caption{\small Branching ratios
 for  Higgs-boson decays into
vector bosons or photons $H\to WW,ZZ, \gamma\gamma,Z\gamma$, and fermions $H\to bb, cc, \tau\tau$  versus $\Lambda$, for 
$m_H= 120$ GeV (left) and 140 GeV (right). Also shown is the branching ratio for the FC decay $H\to bs$.
 }
\label{fig2}
\end{figure}

In Fig.\ref{fig2}, the BR's for the main Higgs-boson decays into
vector bosons and photons $H\to WW,ZZ, \gamma\gamma,Z\gamma$ and fermions $H\to bb, cc, \tau\tau$ are shown versus $\Lambda$, for 
$m_H= 120$ GeV (left) and 140 GeV (right). Also shown is BR($H\to bs$) that will be discussed in Sec. 5.
The enhancement of the decays into vector bosons and photons
is remarkable (see also plots on the corresponding ratios BR/BR${_{SM}}$ in \cite{our}). This is  clearly a bonus for Higgs-boson searches at the LHC. On the other hand, all the branching ratios BR($H\to WW,ZZ, \gamma\gamma,Z\gamma$) are almost   insensitive to the scale $\Lambda$. On the contrary, the Higgs decays into fermions, although generally depleted with respect to their SM rates, show a nice sensitivity to $\Lambda$, and can provide a handle for a possible $\Lambda$ determination. To this respect LHC can hardly contribute, while we will discuss in the next section how  a linear collider could allow  a $\Lambda$ measurement through the direct detection of 
Higgs-boson decays into fermions.

Note that neither EW precision tests nor FC neutral current processes presently constrain the effective-Yukawa scenario \cite{our}.
Also, the experimental exclusion limits on $m_H$ as elaborated in the SM in direct searches \cite{LEP,Tevatron} 
should be revisited in the light of a possible fermionic-coupling depletion that differs from the purely FP limit. A dedicated analysis is needed to obtain $m_H$ bounds in the effective-Yukawa model.  A relaxed direct lower bound on $m_H$  is anyway expected with respect to the SM limit of 114.4 GeV \cite{our}.


\section{Production cross sections for different Higgs boson signatures}

It is well-known that, to a great extent, the precision study of  light-Higgs-boson properties at  linear colliders does not require  running at very high c.m. energies \cite{batt,djou}. Production cross sections are somewhat optimized for
 collision energies $\sqrt{S}$ not much larger than the kinematical threshold for the associated production $e^+e^-\to ZH$. While
the vector-boson-fusion production rate increases as  $\log{S}$, and gets comparable to the cross section for 
 $e^+e^-\to ZH$ (that scales as $1/S$) at  energies 
$\sqrt{S}\sim 500$ GeV, for lower $\sqrt{S}$ the associated production $e^+e^-\to ZH$ has 
the dominant cross section. In particular, for  $ m_H\simeq$ 120 GeV, 
$\sigma(e^+e^-\to Z H)\simeq 0.13$ pb  at 
$\sqrt{S}\simeq 350$ GeV,  to be compared with the corresponding $\sigma(e^+e^-\to\nu\nu H)\simeq 0.03$ 
pb. At $\sqrt{S}\simeq 800$ GeV, $\sigma(e^+e^-\to\nu\nu H)$  increases and gets dominant, but the total production rate is  
$\sigma(e^+e^-\to Z H + \nu\nu H)\simeq (0.02+0.17)$ pb, that is just slightly larger than its value at $\sqrt{S}\simeq 350$ GeV.
The associated production  benefits from the further advantage of the simpler two-body kinematics giving rise (at leading order) to a monochromatic Higgs boson, with  an excellent potential  even in  case of an invisible Higgs boson
\cite{LC}. 

\begin{figure}[tpb]
\begin{center}
\dofigs{3.1in}{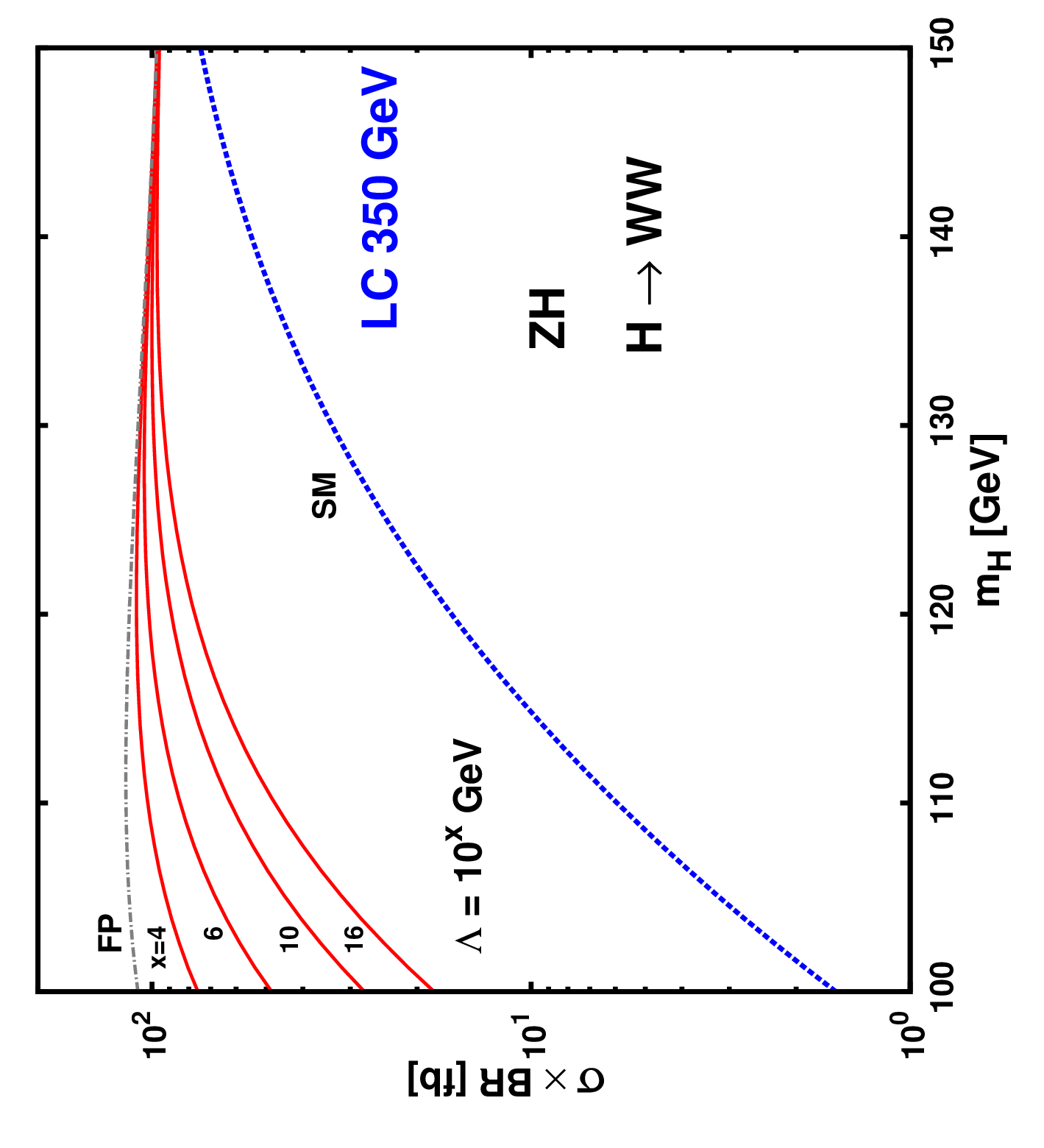}{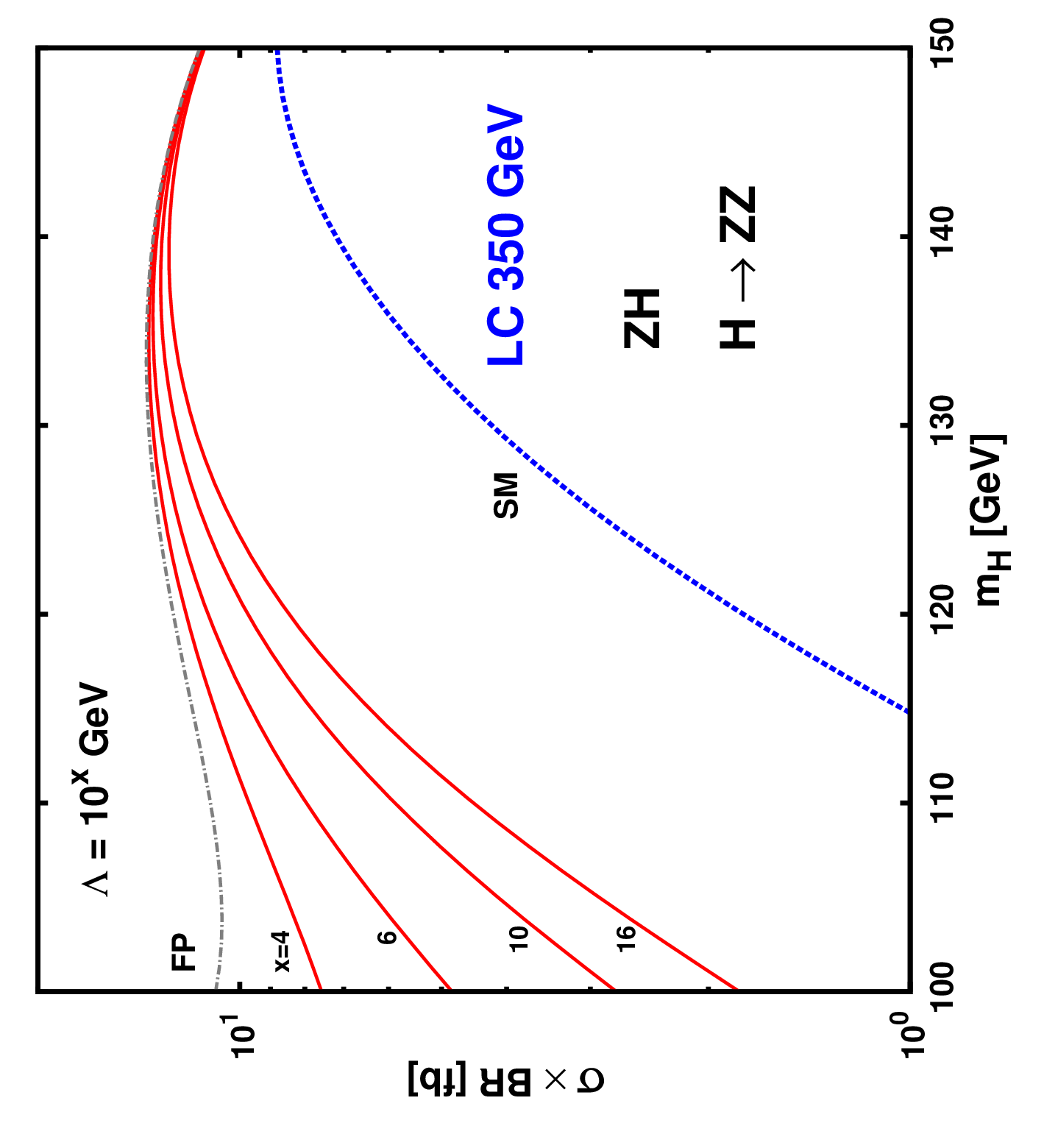}
\end{center}
\caption{\small 
Production cross sections for  $e^+e^-\to Z H$  at $\sqrt{S}\simeq 350$ GeV times the branching ratio for  $H\to
 WW$ (left) and $ZZ$ (right) versus $m_H$,
at different values of $\Lambda$. 
The curves labeled by SM and FB correspond to the standard-model and the naive fermiophobic Higgs scenarios, respectively.
 }
\label{fig3}
\end{figure}

\begin{figure}[tpb]
\begin{center}
\dofigs{3.1in}{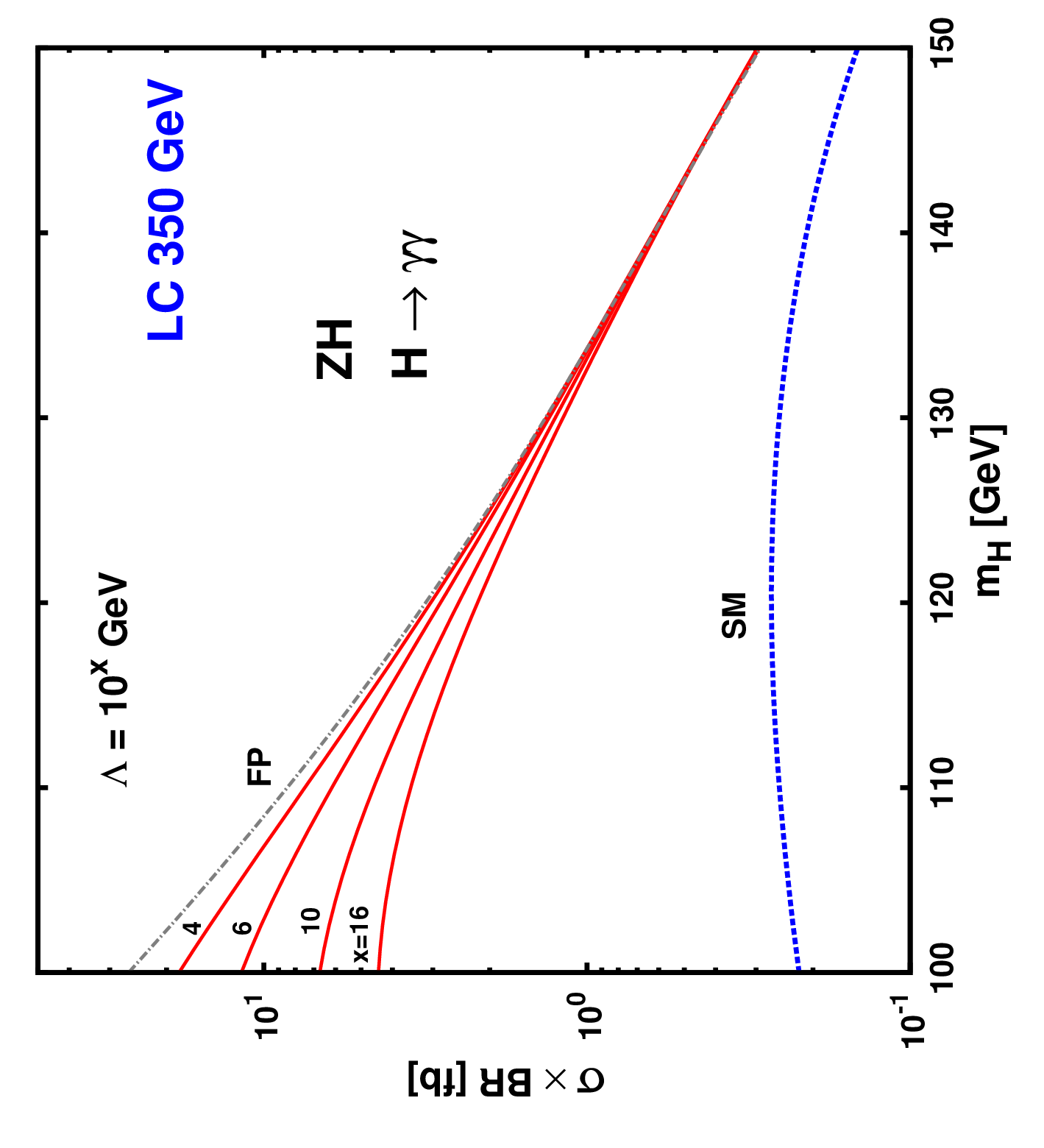}{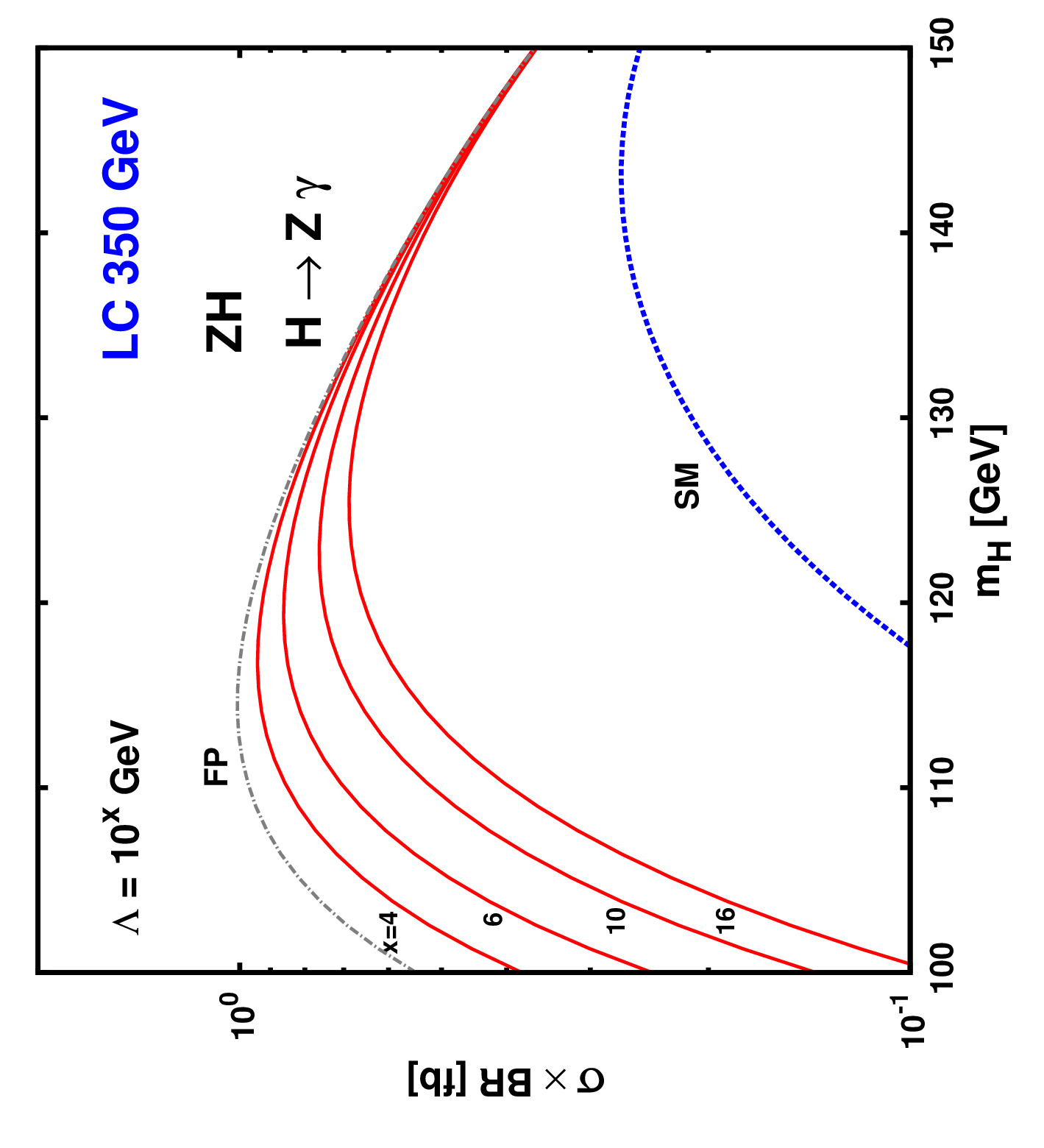}
\end{center}
\caption{\small 
Production cross sections for  $e^+e^-\to Z H$  at $\sqrt{S}\simeq 350$ GeV times the branching ratio for  $H\to
 \gamma\gamma$ (left) and $Z\gamma$ (right) versus $m_H$,
at different values of $\Lambda$. 
The curves labeled by SM and FB correspond to the standard-model and the naive fermiophobic Higgs scenarios, respectively.
 }
\label{fig4}
\end{figure}

\begin{figure}[tpb]
\begin{center}
\dofigs{3.1in}{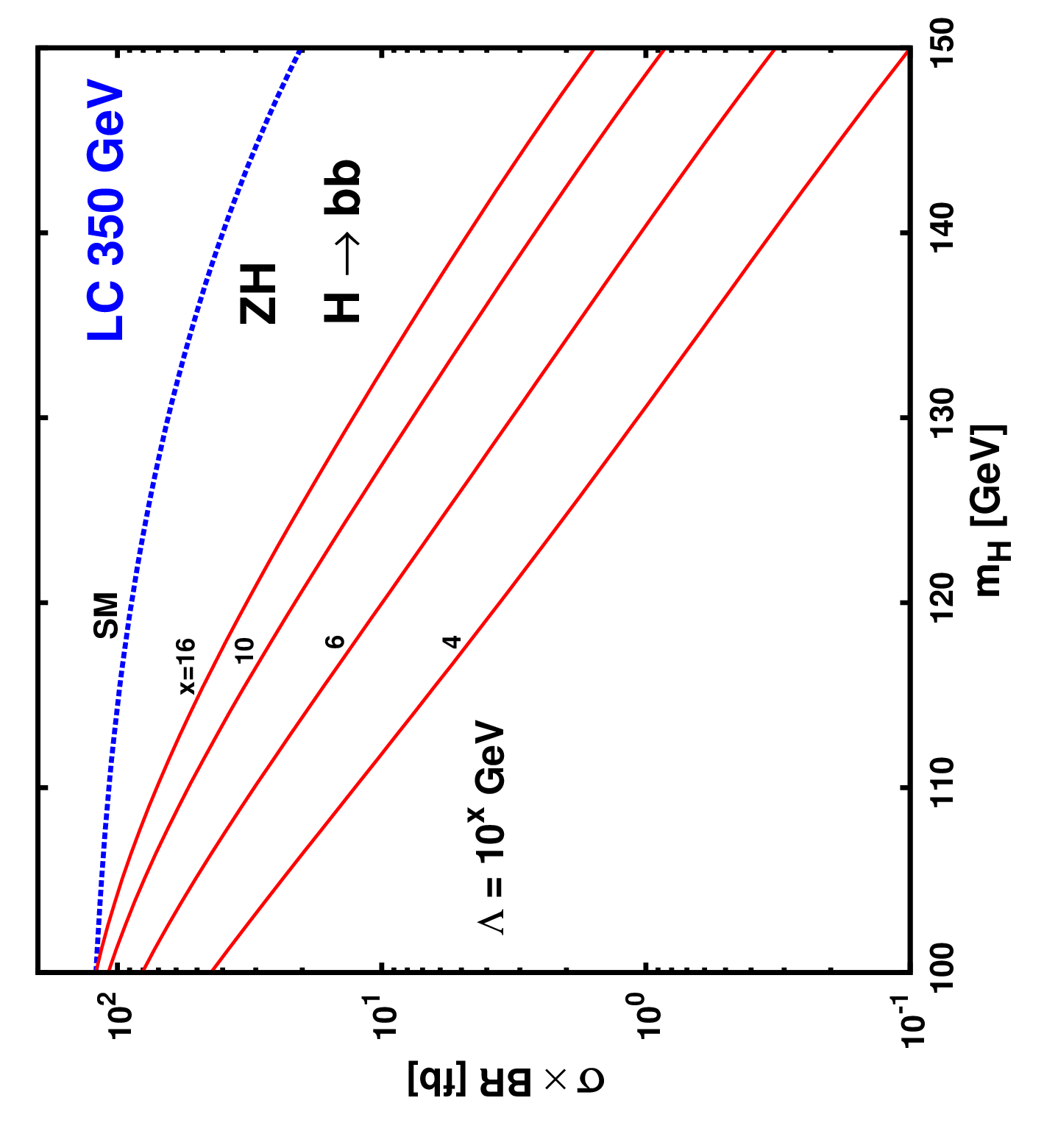}{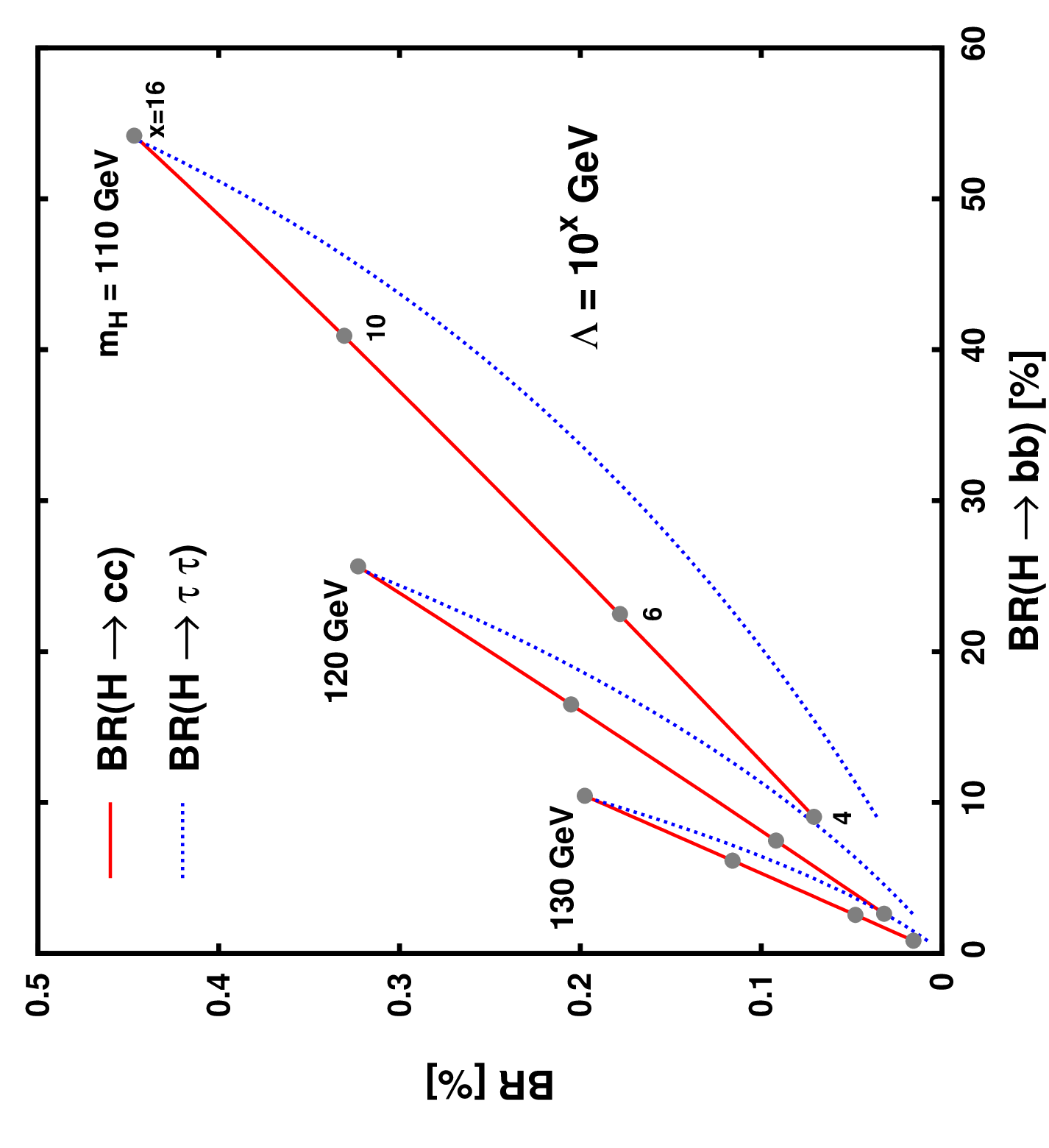}
\end{center}
\caption{\small Left : 
Production cross sections for  $e^+e^-\to Z H$  at $\sqrt{S}\simeq 350$ GeV times the branching ratio for  $H\to
 b\bar b$ versus $m_H$,
at different values of $\Lambda$.  The curves labeled by SM and FB correspond to the standard-model and the naive fermiophobic Higgs scenarios, respectively. Right : Correlation  between the dominant fermionic BR's for different  $m_H$ values.
For each $m_H$ choice, the  $\Lambda$ values are univocally set by  BR$(H\to b\bar b)$ (see text).}
\label{fig5}
\end{figure}

On this basis, we present here the production rates 
for the dominant Higgs boson decays, 
for a linear collider running at $\sqrt{S}\simeq 350$ GeV 
(that allows  top-quark pair production, too). In particular, in Figs.~\ref{fig3}--\ref{fig5}, we plot
the quantities 
$\sigma(e^+e^-\to Z H)\times$BR$(H\to
 WW,ZZ, \gamma\gamma,Z\gamma, b\bar b)$ versus $m_H$,
for different values of $\Lambda$.

Note that the {\it inclusive} Higgs production cross sections are  model ($\Lambda$) independent.
Production rates for a different value of $\sqrt{S}$ and/or for the vector-boson-fusion channel
can be obtained from Figs.~\ref{fig3}--\ref{fig5} by  just rescaling the corresponding cross sections
in the SM.

The typical integrated luminosity collected at linear colliders is expected to be a few hundreds  of fb$^{-1}$, and we present in Table~\ref{table2} the number of expected events $N_{\rm ev}(X)$ 
corresponding to an integrated luminosity of 500 fb$^{-1}$ 
for the production channel  
$e^+e^-\to Z H$, at $\sqrt{S}\simeq 350$ GeV.
Different $H\to X$ decay signatures are considered,  versus 
the Higgs boson  mass  and  $\Lambda$ (both in  GeV units). 
The lower-rate decays $H\to c\bar c$ and $H\to \tau\tau$ are included in Table~\ref{table2}, too.

\begin{table}[htbp]
\begin{center}
                                                                                                                                                                                                                                            \begin{tabular}{|c|c|c|c|c|c|c|c|c|}                             \hline {\large $m_H$} & {$\Lambda$} &  $N_{\rm ev}(\gamma\gamma)$  & $N_{\rm ev}(WW)$ &  $N_{\rm ev}(ZZ)$ & $N_{\rm ev}(Z\gamma)$ & $N_{\rm ev}(b\bar{b})$ & $N_{\rm ev}(c\bar{c})$ &                 $N_{\rm ev}(\tau\bar{\tau})$                                      \\ \hline \hline \multirow{4}{*}{100} & $10^4$ &
   9.1$ \, \cdot\,  10^{3}$                                   &
   3.8$ \, \cdot\, 10^{4}$                                   &
   3.8$ \, \cdot\,  10^{3}$                                   &
   1.9$ \, \cdot\,  10^{2}$                                   &
   2.2$ \, \cdot\, 10^{4}$                                   &
   1.1$ \, \cdot\,  10^{2}$                                   &
   5.6$ \, \cdot\,  10^{1}$                                  \\
     & $10^6$ &
   5.9$ \, \cdot\,  10^{3}$                                   &
   2.4$ \, \cdot\, 10^{4}$                                   &
   2.4$ \, \cdot\,  10^{3}$                                   &
   1.2$ \, \cdot\,  10^{2}$                                   &
   4.0$ \, \cdot\, 10^{4}$                                   &
   2.0$ \, \cdot\,  10^{2}$                                   &
   1.2$ \, \cdot\,  10^{2}$                                  \\
  & $10^{10}$ &
   3.3$ \, \cdot\,  10^{3}$                                   &
   1.4$ \, \cdot\, 10^{4}$                                   &
   1.4$ \, \cdot\,  10^{3}$                                   &
   6.9$ \, \cdot\,  10^{1}$                                   &
   5.4$ \, \cdot\, 10^{4}$                                   &
   2.8$ \, \cdot\,  10^{2}$                                   &
   2.2$ \, \cdot\,  10^{2}$                                  \\
  & $10^{16}$ &
   2.2$ \, \cdot\,  10^{3}$                                   &
   9.0$ \, \cdot\,  10^{3}$                                   &
   9.0$ \, \cdot\,  10^{2}$                                   &
   4.5$ \, \cdot\,  10^{1}$                                   &
   6.0$ \, \cdot\, 10^{4}$                                   &
   3.2$ \, \cdot\,  10^{2}$                                   &
   3.2$ \, \cdot\,  10^{2}$          \\ \hline \multirow{2}{*}{  100     }
         & FP &
   1.3$ \, \cdot\, 10^{4}$                                   &
   5.4$ \, \cdot\, 10^{4}$                                   &
   5.4$ \, \cdot\,  10^{3}$                                   &
   2.7$ \, \cdot\,  10^{2}$                                   &
         0          &
         0          &
         0         \\
         & SM &
   1.1$ \, \cdot\,  10^{2}$                                   &
   7.8$ \, \cdot\,  10^{2}$                                   &
   7.8$ \, \cdot\,  10^{1}$                                   &
         3.5                                   &
   6.0$ \, \cdot\, 10^{4}$                                   &
   2.8$ \, \cdot\,  10^{3}$                                   &
   6.1$ \, \cdot\,  10^{3}$                                  \\
                                                                                                                                                                                                                                                                                                                                                                                                                                                                                                                                                                          \hline \hline \multirow{4}{*}{110}  & $10^4$ &
   3.7$ \, \cdot\,  10^{3}$                                   &
   5.4$ \, \cdot\, 10^{4}$                                   &
   4.9$ \, \cdot\,  10^{3}$                                   &
   5.0$ \, \cdot\,  10^{2}$                                   &
   6.3$ \, \cdot\,  10^{3}$                                   &
   4.9$ \, \cdot\,  10^{1}$                                   &
   2.5$ \, \cdot\,  10^{1}$                                  \\
     & $10^6$ &
   3.2$ \, \cdot\,  10^{3}$                                   &
   4.6$ \, \cdot\, 10^{4}$                                   &
   4.1$ \, \cdot\,  10^{3}$                                   &
   4.3$ \, \cdot\,  10^{2}$                                   &
   1.6$ \, \cdot\, 10^{4}$                                   &
   1.2$ \, \cdot\,  10^{2}$                                   &
   7.7$ \, \cdot\,  10^{1}$                                  \\
  & $10^{10}$ &
   2.4$ \, \cdot\,  10^{3}$                                   &
   3.5$ \, \cdot\, 10^{4}$                                   &
   3.1$ \, \cdot\,  10^{3}$                                   &
   3.2$ \, \cdot\,  10^{2}$                                   &
   2.9$ \, \cdot\, 10^{4}$                                   &
   2.3$ \, \cdot\,  10^{2}$                                   &
   1.8$ \, \cdot\,  10^{2}$                                  \\
  & $10^{16}$ &
   1.9$ \, \cdot\,  10^{3}$                                   &
   2.7$ \, \cdot\, 10^{4}$                                   &
   2.4$ \, \cdot\,  10^{3}$                                   &
   2.5$ \, \cdot\,  10^{2}$                                   &
   3.8$ \, \cdot\, 10^{4}$                                   &
   3.1$ \, \cdot\,  10^{2}$                                   &
   3.1$ \, \cdot\,  10^{2}$          \\ \hline \multirow{2}{*}{  110     }
         & FP &
   4.1$ \, \cdot\,  10^{3}$                                   &
   6.0$ \, \cdot\, 10^{4}$                                   &
   5.4$ \, \cdot\,  10^{3}$                                   &
   5.5$ \, \cdot\,  10^{2}$                                   &
         0          &
         0          &
         0         \\
         & SM &
   1.3$ \, \cdot\,  10^{2}$                                   &
   3.2$ \, \cdot\,  10^{3}$                                   &
   2.9$ \, \cdot\,  10^{2}$                                   &
   2.6$ \, \cdot\,  10^{1}$                                   &
   5.5$ \, \cdot\, 10^{4}$                                   &
   2.5$ \, \cdot\,  10^{3}$                                   &
   5.5$ \, \cdot\,  10^{3}$                                  \\
                                                                                                                                                                                                                                                                                                                                                                                                                                                                                                                                                                          \hline \hline \multirow{4}{*}{120}  & $10^4$ &
   1.5$ \, \cdot\,  10^{3}$                                   &
   5.6$ \, \cdot\, 10^{4}$                                   &
   6.2$ \, \cdot\,  10^{3}$                                   &
   5.0$ \, \cdot\,  10^{2}$                                   &
   1.7$ \, \cdot\,  10^{3}$                                   &
   2.1$ \, \cdot\,  10^{1}$                                   &
   1.1$ \, \cdot\,  10^{1}$                                  \\
     & $10^6$ &
   1.4$ \, \cdot\,  10^{3}$                                   &
   5.3$ \, \cdot\, 10^{4}$                                   &
   5.9$ \, \cdot\,  10^{3}$                                   &
   4.7$ \, \cdot\,  10^{2}$                                   &
   5.0$ \, \cdot\,  10^{3}$                                   &
   6.1$ \, \cdot\,  10^{1}$                                   &
   3.7$ \, \cdot\,  10^{1}$                                  \\
  & $10^{10}$ &
   1.3$ \, \cdot\,  10^{3}$                                   &
   4.8$ \, \cdot\, 10^{4}$                                   &
   5.3$ \, \cdot\,  10^{3}$                                   &
   4.3$ \, \cdot\,  10^{2}$                                   &
   1.1$ \, \cdot\, 10^{4}$                                   &
   1.4$ \, \cdot\,  10^{2}$                                   &
   1.1$ \, \cdot\,  10^{2}$                                  \\
  & $10^{16}$ &
   1.1$ \, \cdot\,  10^{3}$                                   &
   4.3$ \, \cdot\, 10^{4}$                                   &
   4.7$ \, \cdot\,  10^{3}$                                   &
   3.8$ \, \cdot\,  10^{2}$                                   &
   1.7$ \, \cdot\, 10^{4}$                                   &
   2.1$ \, \cdot\,  10^{2}$                                   &
   2.2$ \, \cdot\,  10^{2}$          \\ \hline \multirow{2}{*}{  120     }
         & FP &
   1.5$ \, \cdot\,  10^{3}$                                   &
   5.8$ \, \cdot\, 10^{4}$                                   &
   6.4$ \, \cdot\,  10^{3}$                                   &
   5.1$ \, \cdot\,  10^{2}$                                   &
         0          &
         0          &
         0         \\
         & SM &
   1.4$ \, \cdot\,  10^{2}$                                   &
   8.9$ \, \cdot\,  10^{3}$                                   &
   9.9$ \, \cdot\,  10^{2}$                                   &
   7.0$ \, \cdot\,  10^{1}$                                   &
   4.6$ \, \cdot\, 10^{4}$                                   &
   2.1$ \, \cdot\,  10^{3}$                                   &
   4.6$ \, \cdot\,  10^{3}$                                  \\
                                                                                                                                                                                                                                                                                                                                                                                                                                                                                                                                                                          \hline \hline \multirow{4}{*}{130}  & $10^4$ &
   6.5$ \, \cdot\,  10^{2}$                                   &
   5.4$ \, \cdot\, 10^{4}$                                   &
   7.0$ \, \cdot\,  10^{3}$                                   &
   3.9$ \, \cdot\,  10^{2}$                                   &
   5.3$ \, \cdot\,  10^{2}$                                   &
         9.9                                   &
         5.0                                  \\
     & $10^6$ &
   6.5$ \, \cdot\,  10^{2}$                                   &
   5.3$ \, \cdot\, 10^{4}$                                   &
   6.9$ \, \cdot\,  10^{3}$                                   &
   3.9$ \, \cdot\,  10^{2}$                                   &
   1.6$ \, \cdot\,  10^{3}$                                   &
   3.0$ \, \cdot\,  10^{1}$                                   &
   1.8$ \, \cdot\,  10^{1}$                                  \\
  & $10^{10}$ &
   6.3$ \, \cdot\,  10^{2}$                                   &
   5.1$ \, \cdot\, 10^{4}$                                   &
   6.6$ \, \cdot\,  10^{3}$                                   &
   3.7$ \, \cdot\,  10^{2}$                                   &
   3.8$ \, \cdot\,  10^{3}$                                   &
   7.2$ \, \cdot\,  10^{1}$                                   &
   5.8$ \, \cdot\,  10^{1}$                                  \\
  & $10^{16}$ &
   6.0$ \, \cdot\,  10^{2}$                                   &
   4.8$ \, \cdot\, 10^{4}$                                   &
   6.3$ \, \cdot\,  10^{3}$                                   &
   3.5$ \, \cdot\,  10^{2}$                                   &
   6.5$ \, \cdot\,  10^{3}$                                   &
   1.2$ \, \cdot\,  10^{2}$                                   &
   1.2$ \, \cdot\,  10^{2}$          \\ \hline \multirow{2}{*}{  130     }
         & FP &
   6.5$ \, \cdot\,  10^{2}$                                   &
   5.4$ \, \cdot\, 10^{4}$                                   &
   7.1$ \, \cdot\,  10^{3}$                                   &
   4.0$ \, \cdot\,  10^{2}$                                   &
         0          &
         0          &
         0         \\
         & SM &
   1.3$ \, \cdot\,  10^{2}$                                   &
   1.8$ \, \cdot\, 10^{4}$                                   &
   2.4$ \, \cdot\,  10^{3}$                                   &
   1.2$ \, \cdot\,  10^{2}$                                   &
   3.4$ \, \cdot\, 10^{4}$                                   &
   1.6$ \, \cdot\,  10^{3}$                                   &
   3.4$ \, \cdot\,  10^{3}$                                  \\
                                                                                                                                                                                                                                                                                                                                                                                                                                                                                                                                                                          \hline \hline \multirow{4}{*}{140}  & $10^4$ &
   3.1$ \, \cdot\,  10^{2}$                                   &
   5.1$ \, \cdot\, 10^{4}$                                   &
   6.9$ \, \cdot\,  10^{3}$                                   &
   2.8$ \, \cdot\,  10^{2}$                                   &
   1.7$ \, \cdot\,  10^{2}$                                   &
         4.8                                   &
         2.4                                  \\
     & $10^6$ &
   3.1$ \, \cdot\,  10^{2}$                                   &
   5.1$ \, \cdot\, 10^{4}$                                   &
   6.9$ \, \cdot\,  10^{3}$                                   &
   2.8$ \, \cdot\,  10^{2}$                                   &
   5.3$ \, \cdot\,  10^{2}$                                   &
   1.5$ \, \cdot\,  10^{1}$                                   &
         9.2                                  \\
  & $10^{10}$ &
   3.1$ \, \cdot\,  10^{2}$                                   &
   5.0$ \, \cdot\, 10^{4}$                                   &
   6.8$ \, \cdot\,  10^{3}$                                   &
   2.8$ \, \cdot\,  10^{2}$                                   &
   1.3$ \, \cdot\,  10^{3}$                                   &
   3.8$ \, \cdot\,  10^{1}$                                   &
   3.0$ \, \cdot\,  10^{1}$                                  \\
  & $10^{16}$ &
   3.1$ \, \cdot\,  10^{2}$                                   &
   4.9$ \, \cdot\, 10^{4}$                                   &
   6.6$ \, \cdot\,  10^{3}$                                   &
   2.7$ \, \cdot\,  10^{2}$                                   &
   2.4$ \, \cdot\,  10^{3}$                                   &
   6.7$ \, \cdot\,  10^{1}$                                   &
   6.8$ \, \cdot\,  10^{1}$          \\ \hline \multirow{2}{*}{  140     }
         & FP &
   3.1$ \, \cdot\,  10^{2}$                                   &
   5.1$ \, \cdot\, 10^{4}$                                   &
   7.0$ \, \cdot\,  10^{3}$                                   &
   2.8$ \, \cdot\,  10^{2}$                                   &
         0          &
         0          &
         0         \\
         & SM &
   1.1$ \, \cdot\,  10^{2}$                                   &
   2.8$ \, \cdot\, 10^{4}$                                   &
   3.9$ \, \cdot\,  10^{3}$                                   &
   1.4$ \, \cdot\,  10^{2}$                                   &
   2.1$ \, \cdot\, 10^{4}$                                   &
   9.6$ \, \cdot\,  10^{2}$                                   &
   2.1$ \, \cdot\,  10^{3}$             \\ \hline \end{tabular}

\end{center} 
\caption[]{Number of expected events $N_{\rm ev}(X)$ 
for
an integrated luminosity of 500 fb$^{-1}$, corresponding to 
$e^+e^-\to Z H\to Z X$ at $\sqrt{S}\simeq 350$ GeV, for different
Higgs-boson decays $H\to X$, versus 
the Higgs-boson  mass $m_H$  and  $\Lambda$ (both in  GeV units). 
The SM and FP labels stand for 
the  standard-model  and  the fermiophobic Higgs scenario results, respectively.}
\label{table2} 
\end{table}
Figure \ref{fig3} shows  production rates for 
$H\to WW$ (left) and $ZZ$ (right). Cross sections are quite enhanced with respect to the SM at low $m_H$. They  are large enough to allow 
an accurate study of both channels, by exploiting both the leptonic and the hadronic $W/Z$ decays. For instance, at $m_H \simeq 110$ GeV, for $\Lambda= (10^{4}$ to $10^{16})$ GeV, one expects (5.4 to $2.7) \times 10^4$ $WWZ$ events 
(to be compared with $3.2\times 10^3$ in the SM), and (4.9 to $2.4) \times 10^3$ $ZZZ$ events (to be compared with $2.9\times 10^2$ in the SM) (cf. Table~\ref{table2}).  At lower $m_H$, the sensitivity to $\Lambda$ increases, while at larger $m_H$ a $\Lambda$ determination becomes more and more difficult.

A similar pattern, as far as both rate enhancement  and sensitivity to $\Lambda$ are concerned, is found for the $H\to \gamma\gamma$ channel [cf. Fig.~\ref{fig4} (left)], that is anyway characterized by a cleaner signature (a $\gamma\gamma$ resonance). 
In particular, for $m_H \simeq 110$ GeV, and $\Lambda= (10^{4}$ to $10^{16})$ GeV, one expects  
(3.7 to $1.9) \times 10^3$ $Z\gamma\gamma$ events (to be compared with $1.3\times 10^2$ in the SM) (cf. Table~\ref{table2}). 
Lower rates are predicted for $H\to Z\gamma$ [cf. Fig.~\ref{fig4} (right)], for which, anyway, a few hundreds of  events
are expected in most of the parameter space.

In Fig.~\ref{fig5} (left), the production rates for the $H\to b\bar b$ decay channel are shown.  The $H\to b\bar b$ channel gives a remarkable opportunity to make an accurate $\Lambda$ determination in all the $m_H$ range considered here.
Not only the $H\to b\bar b$ rate is quite sensitive to $\Lambda$ at low 
$m_H$, but this sensitivity even increases at high $m_H$'s
(unlike what occurs for the $H\to WW,ZZ,\gamma\gamma,Z\gamma$ channels).
 At $m_H \simeq 110$ GeV,
 (6.3 to $38) \times 10^3$ $b\bar b$ events are predicted (to be compared with $5.5\times 10^4$ in the SM), for $\Lambda= (10^{4}$ to $10^{16})$ GeV (cf. Table~\ref{table2}).
  At $m_H \simeq 140$ GeV, the rate is lower but the sensitivity to $\Lambda $ is much larger. In particular, for $\Lambda= (10^{4}$ to $10^{16})$ GeV, one expects
  (1.7 to $24) \times 10^2$ $b\bar b$ events  (to be compared with $2.1\times 10^4$ in the SM).
 
  The numbers of events corresponding to the channels  
  $H\to c\bar c$ and $H\to \tau\tau$ are quite suppressed with respect to the SM values.  Anyway,
   the few tens or hundreds of  events
 expected in most of the parameter space (cf. Table~\ref{table2}) should allow a fair BR's determination for the corresponding decays.
 
 In Fig.\ref{fig5} (right), we show the  correlations between the BR's for the decays $H\to c\bar c$ and $H\to \tau\tau$, and  BR$(H\to b\bar b)$,  for 
 $m_H=$110, 120, 130 GeV. For each $m_H$ value,  $\Lambda$ is univocally set by  
BR$(H\to b\bar b)$, and we report the points corresponding to $\Lambda= 10^{4}, 10^{6},10^{10},10^{16}$ GeV (grey bubbles) only on the $H\to c\bar c$ curves (related points on the $H\to \tau\tau$
curves can be easily inferred). These correlations are characteristics of the radiative structures of the Yukawa-coupling generation.
BR$(H\to c\bar c)$ depends linearly on BR$(H\to b\bar b)$, reflecting a similar structure of the corresponding RG equations.
Nonlinear differences in the behavior of BR$(H\to\tau\tau)$ arise from the different impact of  strong interactions  on the 
leptonic-coupling evolution with respect to the quark case (see Section 4).

The rates for the different decay channels
in Table~\ref{table2} 
can  give a first hint on how accuracies in the measurement 
of various Higgs-boson couplings could scale with respect to the corresponding SM values.  
In previous  Higgs-boson studies \cite{LC,Battaglia:1999re}, the expectations for the precision on the Higgs  branching ratios and 
couplings have been reported  for  linear colliders with $\sqrt{S}\simeq 350$ GeV and 500 GeV and  integrated luminosity of the order of 500
fb$^{-1}$. 
A similar precision is then expected for the setup assumed in Table~\ref{table2}.
The  relative precision  on the measurements of the SM branching ratios  
BR($H\to b\bar b,c\bar c, \tau\tau,WW,ZZ$)
 is a few percent for $m_H\simeq 120$ GeV 
 \cite{LC,Battaglia:1999re}. The accuracy on BR$(H\to \gamma\gamma)$ is a bit lower \cite{LC}.
In case the effective Yukawa scenario is realized, accuracies on the measurements of 
BR($H\to WW,ZZ,\gamma\gamma,\gamma Z$) will be much better than in the SM.
The precision  on the measurement of 
BR$(H\to b\bar b)$
 will be  comparable with the SM estimate 
  at very low  $m_H$, while getting worse in the intermediate and large   $m_H$ range.
  On the other hand, accuracies on BR$(H\to c\bar c)$ and especially on 
 BR$(H\to \tau\tau)$ are expected to deteriorate with respect to the SM case in all the $m_H$ range.
 A more quantitative analysis would require going into the relevant backgrounds and detection efficiencies.

\section{Effective flavor-changing Yukawa couplings}

In this section we  analyze the flavor-changing (FC) fermionic 
 decays
\bea
H \to f_i f_j \equiv \bar{f}_i f_j+\bar{f}_j f_i
\label{processFC}
\eea
where the $i\neq j$ indices stand for  generic  flavors,  in the up-quark (or 
down-quark) sectors.
In the SM, the decay amplitudes for $H \to f_i f_j$
are generated at one loop,  
and are finite, thanks to the unitarity of the CKM matrix. These decays are characterized by  very small BR's.
Even the decay $H\to bs$, that is not suppressed by the 
Glashow-Iliopoulos-Maiani (GIM)  mechanism because of 
 the unbalanced top-quark contribution in the loop, has a quite small BR.
In particular, for $m_H < 2 M_W$,
one has BR$(H\to b s) \simeq 2 \cdot 10^{-7}$ \cite{Eilam,Arhrib}, which makes this channel practically undetectable in the SM.

On the other hand, the small BR$(H\to b s)$ makes  
the Higgs decay  $H\to b s$ 
a  sensitive probe of  
potential new physics contributions above the EW scale.
This process has been extensively considered in literature, 
with emphasis on minimal and non-minimal supersymmetric extensions
of the SM \cite{Arhrib,HbsSUSY}, where the corresponding  
BR$(H\to b s)$ can be 
as large as ($10^{-4}-10^{-3}$)  in particular configurations of the allowed SUSY parameter space.

In the following, we will compute BR$(H\to b s)$ in the effective Yukawa scenario, and find that it can also be 
in the range  $(10^{-4}-10^{-3})$ for $m_H \lsim 140$ GeV.

In case a new  mechanism for ChSB and generation of 
fermion masses exists,  
it is natural to assume that it will generate
a fermion mass matrix on the fermion weak eigenstates which is equal to the SM one.
 The CKM is then obtained as usual by rotating the fermion fields into the fermion mass eigenstates.

Fermion masses 
explicitly breaking  chiral symmetry 
 radiatively induces both  flavor-diagonal and 
flavor-changing Yukawa couplings because of  
the off-diagonal terms in the CKM matrix.
Then for a light Higgs boson one gets  a large enhancement in the FC Higgs decay BR's  arising from two combined effects. 
On the one hand, the Higgs total width  is depleted with respect to the
SM one, being the $b$-quark Yukawa coupling radiatively generated. 
On the other hand, there is a significant effect in the resummation of the leading log terms 
for the FC amplitude for $\Lambda \gg m_H$. 
Moreover,  the ratio of  the FC decay amplitude for  the decay 
$H\to b s$ to the  flavor-conserving  $H\to \bar{b} b$ one will not
be suppressed by gauge couplings and loop factors as in the SM, 
but  will only be  depleted by the  
CKM matrix element $V_{ts}$.
The same holds for  other FC Higgs decays,
although an extra suppression by the GIM mechanism will in 
general affect the ratio. 
Therefore, a large enhancement in the FC Higgs BRs is naturally 
expected in our framework.

In order to calculate  BR$(H\to \bar{f}_i f_j)$ (with $i\neq j$), 
we start by evaluating the effective flavor-changing Yukawa couplings related
to the corresponding $H \bar{f}_i f_j$ interaction term in the Lagrangian.
The FC one-loop $H\to \bar{f}_i f_j$ amplitude is divergent in this scenario, 
unlike in the SM, since tree-level Yukawa couplings are missing. 
In the language of effective field theories, 
this implies that the corresponding FC 
Yukawa coupling $H\bar{f}_i f_j$ 
has to be renormalized at some  high-energy scale.
Then,  Yukawa couplings at low energy can be computed by 
RG equations.

Yukawa couplings, in the fermion mass eigenstates, are defined by the  Lagrangians
for the flavor-conserving interactions, 
\bea
{\cal L}^{\rm Y}_{H}=-\sum_{\it i} \frac{H}{\sqrt{2}} \left( 
{\rm Y_{U_{\it i}}}[\bar{u}_i u_i] + 
 {\rm Y_{D_{\it i}}}[\bar{d}_i d_i]
+ {\rm Y_{E_{\it i}}}[\bar{e}_i e_i]\right) 
\; ,
\label{LY}
\eea
where $i=1,2,3$ for $u_{i}=(u,c,t)$, $d_i=(d,s,b)$, and $e_i=(e,\mu,\tau)$, 
respectively, and the FC interactions\footnote{
In Eq.(\ref{LFCY}) we have not included the contribution of 
FC interactions in the 
charged leptonic sector, that are vanishing in the massless neutrino
limit.}
\bea
{\cal L}^{\rm FCY}_{H}=-\sum_{\it ij} \frac{H}{\sqrt{2}} \left( 
 [\YUL]_{ ij} [\bar{u}_i P_L u_j] + 
[\YUR]_{ij} [\bar{u}_i P_R u_j] +
[\YDL]_{ij} [\bar{d}_{i} P_L d_{j}] + 
[\YDR]_{ij} [\bar{d}_{i} P_R d_{j}]
\right)
\label{LFCY}
\eea
where the indices $i\neq j$ run over the fermion generations, 
$H$ is the Higgs boson field, 
$P_{L/R}=(1\mp \gamma_5)/2$ and, being  ${\cal L}^{\rm FCY}_{H}$ Hermitian, the matrices
$[{\bf Y^{L,R}_{U,D}}]_{ij}$ satisfy 
the condition $({\bf Y^{L}_{U,D}})^{\dag}={\bf Y^{R}_{U,D}}$.
The diagonal entries of ${\bf Y_{U,D}^{L,R}}$ are zero, since
the corresponding flavor-conserving contribution is described by  the flavor-conserving 
Yukawa couplings ${\rm Y_{U_{\it i}}}$, ${\rm Y_{D_{\it i}}}$ in
Eq.(\ref{LY}).  On the other hand, 
left-handed and right-handed two-fermion operators in 
Eq.(\ref{LFCY}) have different radiative couplings whenever 
 initial and final fermions have different masses.
From now on, we will neglect CP violating effects 
in the CKM matrix, and all the Yukawa couplings will be real numbers.
 
We first recall the RG equations for the flavor-conserving 
Yukawa couplings ${\rm Y_{U_{\it i}}}$, ${\rm Y_{D_{\it i}}}$,
${\rm Y_{E_{\it i}}}$. 
In a compact matrix notation, this is given by 
\bea
\frac{d {\bf Y_{F}}}{d t}&=& \beta_{\bf F}\, ,
\label{RGE1}
\eea
where the (diagonal) beta function matrices 
$\beta_{\bf F}$, with ${\scriptstyle {\bf F}}=\{{\bf 
{\scriptstyle U,D,E} } \}$,  are \cite{our}
\bea
\beta_{\bf U}&=&\frac{1}{16 \pi^2}\left\{
3\,\xi_H^2 \left( {\bf Y_U}- {\bf Y^{\scriptscriptstyle SM}_U}\right) - 
3{\bf Y^{\scriptscriptstyle SM}_U}{\bf Y^{\scriptscriptstyle SM}_D}\left({\bf Y_D}-{\bf Y^{\scriptscriptstyle SM}_D}\right)
+\frac{3}{2} {\bf Y_U}\left({\bf Y_U}{\bf Y_U} -
{\bf Y^{\scriptscriptstyle SM}_D} {\bf Y^{\scriptscriptstyle SM}_D}\right)
\right.
\nonumber\\
&-& \left. 
{\bf Y_U}\left(C_{\bf U}\, g_1^2 +\frac{9}{4}g_2^2+8g_3^2-
{\bf Tr(Y)} \right)
\right\}\, , 
\\
\nonumber\\
\beta_{\bf D}&=&\beta_{\bf U}
\{( {\bf {\scriptstyle U,D})}
\rightarrow
({\bf  {\scriptstyle D,U})} \}\, ,
\label{RGE2}\\
\nonumber\\
\beta_{\bf E}&=&\frac{1}{16 \pi^2}\left\{
3\, \xi_H^2 \left( {\bf Y_E}- {\bf Y^{\scriptscriptstyle SM}_E}\right) 
+\frac{3}{2} {\bf Y_E}{\bf Y^{}_E}{\bf Y_E}
-{\bf Y_E}\left(\frac{9}{4}\left( g_1^2 +g_2^2\right)
-{\bf Tr(Y)} \right)
\right\}\, ,
\label{RGE3}
\eea
where $t=\log \mu$, $C_{\bf U}=17/20$, $C_{\bf D}=1/4$,
${\bf Tr(Y)}$ stands for the trace of 
the matrix ${\bf Y }$, and  ${\bf Y }$ is defined as
\bea
{\bf Y }&\equiv &N_c {\bf Y^{}_U}{\bf Y_U}+
N_c {\bf Y^{}_D}{\bf Y_D}+{\bf Y^{}_E}{\bf Y_E}\, .
\label{YYY}
\eea
In particular, ${\bf Y_{U,D,E}}$ (where ${\bf {\scriptstyle U,D,E}}$ 
stand for up-quarks, down-quarks and charged leptons, respectively) are diagonal 
matrices in flavor space, 
${\bf Y_{U,D,E}}={\rm diag}[{\rm Y_{U_{1},D_{1},E_{1}},Y_{U_{2},D_{2},E_{2}},Y_{U_{3},D_{3},E_{3}}}]$. Note that the effective Yukawa couplings for leptons enters  
the effective Yukawa couplings for quarks through Eq.~(\ref{YYY}).
Also, 
\bea
\xi_H\equiv \frac{g_2 m_H}{2 M_W},~~~~~ {\bf Y^{\scriptscriptstyle SM}_F}\equiv
\frac{g_2}{\sqrt{2} M_W} {\rm diag}[ { m_{\rm F_1},m_{\rm F_2},m_{\rm F_3}}],~~~~~
g_1^2\equiv \frac{5}{3}\frac{e^2}{\cos^2{\theta_W}}\, ,
\label{defin}
\eea
where ${\bf Y^{\scriptscriptstyle SM}_F}$ is a diagonal matrix in flavor space,
${ m_{\rm F_{\it i}}}$ being the fermion pole masses,
with ${\scriptstyle {\bf F}}=\{{\bf 
{\scriptstyle U,D,E} } \}$, and $N_c=3$ the number of colors. 
The RG equations for the gauge couplings are the  SM ones \cite{arason}
\bea
\frac{d g_i}{dt}= -b_i \frac{g_i^3}{16 \pi^2}\, ,
\label{RGEg}
\eea
with
$b_1=-\frac{4}{3} n_{g} -\frac{1}{10}\, , ~~~
b_2=\frac{22}{3}-\frac{4}{3} n_{g} -\frac{1}{6}\, , ~~~
b_3=11-\frac{4}{3}n_g\, ,
$
and $n_g=3$ the number of fermion generations.
Terms in ${\bf Y^{\scriptscriptstyle SM}_F}$ give rise to 
ChSB, and are normalized as the tree-level SM 
Yukawa couplings. In deriving Eqs.(\ref{RGE1})-(\ref{RGE3}),
we neglected  subdominant contributions induced by the 
off-diagonal CKM matrix elements in the charged-current weak corrections.
In this approximation, the RG equations for the flavor-diagonal  Yukawa couplings 
do not involve the FC Yukawa couplings.

\begin{figure}[!htb]
\begin{center}
\dofigA{3.1in}{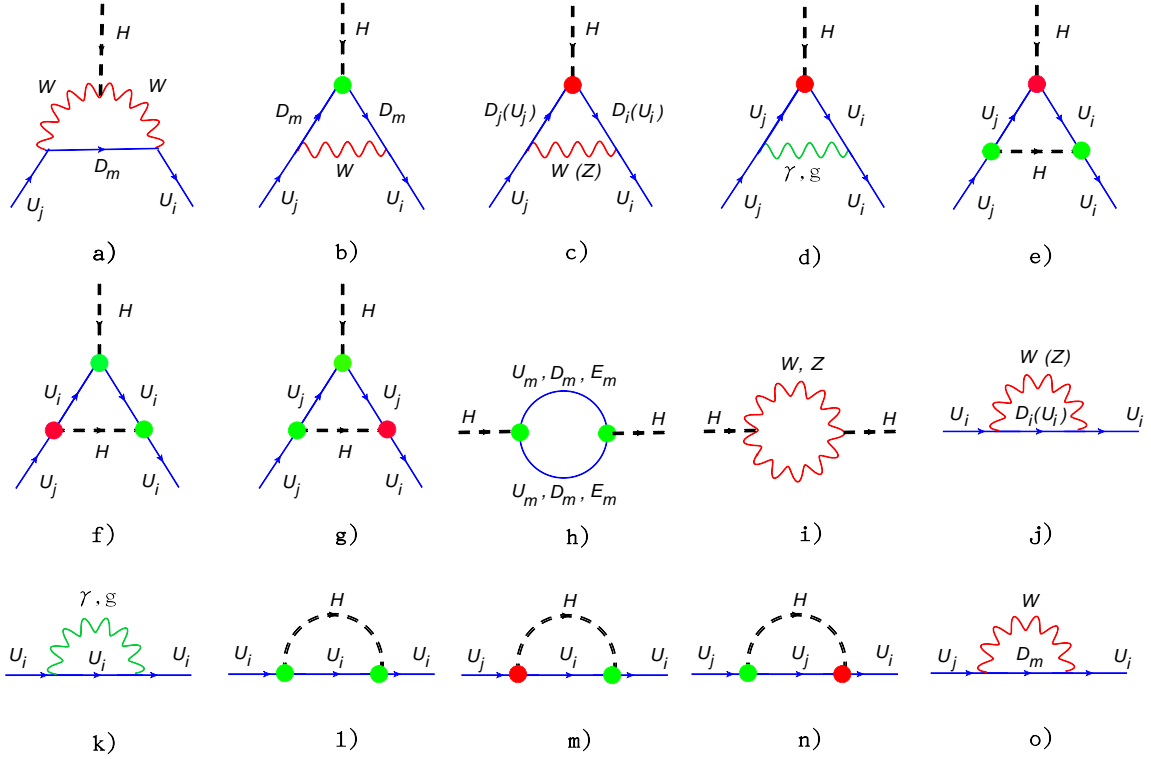}
\end{center}
\caption{\small One-loop Feynman diagrams, in the unitary gauge,   contributing
to the $\beta$ function of the Yukawa matrix elements
$[{\bf Y^L_U}]_{ij}$ and $[{\bf Y^R_U}]_{ij}$ in the up-quark sector, with $U_{i}=\{u,c,t\}$, $D_{i}=\{d,s,b\}$, and 
$E_i=\{e,\mu,\tau\}$, and  $i\neq j$. A sum over the $m$  index 
is understood. Labels
$\gamma$ and $g$  mark the photon and gluon propagators, 
respectively. Diagrams 6(a)-6(g) correspond to the vertex corrections.
Diagrams 6(h)-6(i) contribute to the Higgs boson $H$ self-energy,
while diagrams 6(j)-6(l) and 6(m)-6(o) correspond to the flavor-diagonal and 
FC self-energies in the up-quarks fields sector, respectively. 
Green (light) and red (dark) bubbles represent 
vertex insertions for the flavor-diagonal and FC Yukawa couplings,  respectively. Contributions of double FC vertex insertions have been neglected.
}
\label{diagrams}
\end{figure}

Now we discuss the RG equations for the FC Yukawa couplings
$[{\bf Y^{L}_{U,D}}]_{ij}$ and $[{\bf Y^{R}_{U,D}}]_{ij}$ defined by Eq.(\ref{LFCY}).
Diagrams related to the corresponding $\beta$ functions 
are shown in Fig. \ref{diagrams},
where we have included the full set of EW  
[Fig.6(a)-6(d), 6(i)-6(k), and 6(o)],  strong [Fig.6(d) and 6(k)], and Yukawa 
[Fig.6(e)-6(h) and 6(l)-6(n)] corrections.  In Fig. \ref{diagrams}, 
the green (light) bubbles [in Fig.6(b), 6(e)-6(h), and 6(l)-6(n)]
and red (dark) bubbles [in Fig.6(c)-6(g) and 6(m)-6(n)] 
stand for the vertex insertion of flavor-diagonal and FC Yukawa couplings, 
respectively. Contributions of double FC vertex insertions have been neglected, as discussed further on.

For vanishing tree-level Yukawa couplings,
the leading contribution to the $\beta$ function 
is given by the diagram in Fig.6(a), where two $W$s are exchanged
in the $Hf_if_j$ vertex diagram. Indeed,
the residue at the pole in diagram in Fig.6(a) is the only contribution 
to the Yukawa $\beta$ functions which is not proportional 
to Yukawa couplings.
Then, when all Yukawa couplings  are set to zero 
at the energy scale $\Lambda$, as required by the condition of 
Higgs-fermion decoupling, Yukawa couplings are radiatively 
generated at any energy scale different from $\Lambda$ (here, in particular, at the scale $m_H$) thanks to the  diagram in Fig.6(a). 

By including the full set of corrections  
in Fig.(\ref{diagrams}), we obtain the  RG equations
for the FC Yukawa couplings\footnote{
We stress that the RG equations in  
Eqs.~(\ref{RGE1})-(\ref{RGE3}), and (\ref{betaUL})-(\ref{betaDR})
are also valid in a more general scenario in which the Yukawa couplings 
are not vanishing at tree-level, and  are different from their  
tree-level SM values, provided their tree-level values are small enough  not to 
spoil the perturbative regime.}
\bea
\frac{d [\YFLR]_{ ij}}{dt}&=& [\beta^{{\bf L,R}}_{{\bf F}}]_{ij} \; ,
\label{betaUL}
\eea
where the corresponding beta functions $[\beta^{{\bf L,R}}_{{\bf F}}]_{ij}$,
with  ${\scriptstyle {\bf F}}=\{{\bf {\scriptstyle U,D} } \}$, are given by 
\bea
[\beta^{\bf L}_{\bf U}]_{ij}&=&\frac{1}{16 \pi^2}\left\{
3\sum_{m} \lambda^{ij}_{{\rm D}_m}
\left[{\rm \YSM_{D_{\it m}}} {\rm \YSM_{U_{\it i}}} 
\left({\rm \YSM_{D_{\it m}}}- {\rm Y_{D_{\it m}}}  \right)
+\rho({\scriptstyle {\rm U}_i},{\scriptstyle {\rm U}_j},
{\scriptstyle {\rm D}_m})
{\rm \YSM_{U_{\it j}}}
\left(
{\rm Y_{U_{\it i}}} {\rm \YSM_{U_{\it j}}}-{\rm Y_{U_{\it j}}}{\rm \YSM_{U_{\it i}}}
\right)
\right]
\right.
\nonumber\\
&+& \left.
[\YUL]_{ ij}
\left[3\xi_H^2 -
C_{\bf U}\,  g_1^2 -\frac{9}{4}g_2^2 - 8 g_3^2 +{\bf Tr(Y)}
+\frac{1}{4}\left( {\rm Y^2_{U_{\it i}}}+{\rm Y^2_{U_{\it j}}}\right)
\right. \right.
\nonumber\\
&+& \left.  \left. \frac{3}{2}\left(
{\rm Y^2_{\rm U_{\it j}}}\rho({\scriptstyle {\rm U}_i},
{\scriptstyle {\rm U}_j},{\scriptstyle {\rm U}_i})-
{\rm Y^2_{\rm U_{\it i}}}\rho({\scriptstyle {\rm U}_i},
{\scriptstyle {\rm U}_j},{\scriptstyle {\rm U}_j})\right)
+({\rm \YSM_{U_{\it j}}})^2\left(\frac{3}{2}-\frac{2}{3} s^2_W\right)
\right.\right.
\nonumber\\ 
&+& \left.\left.
({\rm \YSM_{U_{\it i}}})^2\left(-\frac{1}{2}+\frac{2}{3} s^2_W\right)
-\frac{3}{2}({\rm \YSM_{D_{\it j}}})^2
\right]
- 2 [\YDR]_{ij} {\rm \YSM_{U_{\it i}}}{\rm \YSM_{D_{\it j}}}
-  [\YDL]_{ij} {\rm \YSM_{U_{\it i}}}{\rm \YSM_{D_{\it i}}}
\right.
\nonumber\\
&+& \left. [\YUR]_{ij}\left[
\frac{5}{2}{\rm Y_{U_{\it i}}}{\rm Y_{U_{\it j}}}-
{\rm \YSM_{U_{\it i}}}{\rm \YSM_{U_{\it j}}}
-\frac{3}{2}\left({\rm Y^2_{U_{\it i}}}-{\rm Y^2_{U_{\it j}}}\right)
\eta({\scriptstyle {\rm U}_i},{\scriptstyle {\rm U}_j})
\right]
\right\}  \; ,
\eea
\bea
\!\!\!\!\!\!\!\!\!\!\!\! \!\!\!
[\beta^{\bf L}_{\bf D}]_{ij}&=&
[\beta^{\bf L}_{\bf U}]_{ij}
\{( {\bf {\scriptstyle U,D}})
\rightarrow
({\bf  {\scriptstyle D,U}}) \}\, ,  \;\;\;\;\;\;\;\;\;
[\beta^{\bf R}_{\bf U}]_{ij}~=~[\beta^{\bf L}_{\bf U}]^{*}_{ji}\, ,  
\;\;\;\;\;\;\;\;\;
[\beta^{\bf R}_{\bf D}]_{ij}~=~[\beta^{\bf L}_{\bf D}]^{*}_{ji}\, ,  
\;\;\;\;\;\;\;\;\;
\label{betaDR}
\eea
where $\rho({ x},{ y},{ z})=m_{z}^2/(m_{x}^2-m_{y}^2)$,~ 
$\eta({ x},{ y})=m_xm_y/(m_{x}^2-m_{y}^2)$,~
$\lambda^{ij}_{{\rm D}_m}=K_{i m} K^{*}_{ j m}$,
$\lambda^{ij}_{{\rm U}_m}=K_{m i}^{*} K_{m j}$ (with $i\neq j$),
$K_{ij}$ are the CKM matrix elements, and $s_W$
the sine of the Weinberg angle. Since 
$[\beta^{\bf R}_{\bf F}]~=~[\beta^{\bf L}_{\bf F}]^{\dag}$,
right-handed couplings can simply be 
obtained from the left-handed ones
by the general  condition $({\bf Y^{L}_{U,D}})^{\dag}={\bf Y^{R}_{U,D}}$.

In deriving Eqs.(\ref{betaUL})-(\ref{betaDR}),  
we  neglected, in  diagram in Fig.6(c), terms 
of order ${\cal O}(K_{ij}K_{mn})$ (with   $i\neq j$, $m\neq n$),
and related fermion self-energies contributions. 
Indeed, FC couplings  (entering the red bubble)
are naturally of order ${\cal O}(K_{ij})$ in our framework. Then, 
in the $W$-exchange vertices in Fig.6(c)  we kept only diagonal CKM couplings.
Consistently, we neglected also contributions coming from double FC vertex 
insertions.

Notice that in  Eqs. (\ref{betaUL})-(\ref{betaDR}), 
terms that are not proportional to the FC couplings $[{\bf Y^{L,R}_{U,D}}]_{ij}$ 
vanish in the SM limit  ${\rm \YSM_{U_{\it i}}} \to 
{\rm Y_{U_{\it i}}}$.
Indeed, because  of the SM renormalizability, the FC interactions in the  Higgs sector are finite in the SM, implying that the SM $\beta$ functions for the FC couplings  vanish.

In Eqs.(\ref{betaUL})-(\ref{betaDR}), 
we do not find, as we do in Eqs.(\ref{RGE1})-(\ref{RGE3}), any large term
proportional to $\xi_H \propto m_H^2/m_W^2$ multiplied by the ChSB factor   
${\rm \YSM_{U,D}}$.
 Indeed, these terms
are in principle generated by the diagram in Fig.6(a), but their total contribution 
vanishes because of the GIM mechanism and CKM unitarity. On the contrary, the above terms provide the leading contribution 
to the RG equations for the flavor-diagonal Yukawa couplings 
Eqs.~(\ref{RGE1})-(\ref{RGE3}), and are responsible for the breaking of 
perturbative unitarity in the Yukawa sector at large $m_H$ 
\cite{our}.
Notice that contributions proportional to  $\xi_H$   in 
Eqs.(\ref{betaUL})-(\ref{betaDR})  arise from diagrams in Fig.6(c)
and corresponding self-energy contributions [diagrams 6(i)-6(j)]
to the FC vertex corrections, where the GIM mechanism is not active.
On the other hand, 
they are strongly suppressed by the FC $[{\bf Y^{L,R}_{U,D}}]_{ij}$ factors, and 
could endanger perturbative unitarity  only for $m_H$ much larger than the $m_H$ range where the flavor-diagonal equations Eqs.~(\ref{RGE1})-(\ref{RGE3}) are in the perturbative regime \cite{our}.

Following the approach in \cite{our}, our  
renormalization conditions will consist in assuming  all the Yukawa couplings vanishing 
at the scale $\Lambda$, namely 
$$
{\rm Y_{U_{\it i},D_{\it i},E_{\it i}}}(\mu=\Lambda)=0\, , 
\;\;\;\;\;\;\;  [{\bf Y^{L,R}_{U,D}}]_{ij}(\mu=\Lambda)=0\, . 
$$
Then, the corresponding values at  low energy (in particular at $\mu \simeq m_H$)
will be determined by numerically solving the full set of RG equations 
in Eqs.~(\ref{RGE1})-(\ref{RGE3}) and  (\ref{betaUL})-(\ref{betaDR}).

\begin{table}[htbp]
\begin{center}
                                                                                             \begin{tabular}{|c|c|c|c|c|c|c|}                                   \hline { $m_H({\rm GeV})$} &                                      { $\Lambda({\rm GeV})$}                                           & {$|[{\bf Y^L_D}]_{23}|$} &                                      {$|[{\bf Y^R_D}]_{23}|$}  & {$|[{\bf Y^L_U}]_{23}|$ } &           {$|[{\bf Y^R_U}]_{23}|$} &                                        { $|{\rm Y}_b|$}                                                \\ \hline \hline \multirow{4}{*}{100} & $10^4$ &
   1.8$ \, \cdot\, 10^{-6}$                                   &
   8.3$ \, \cdot\, 10^{-5}$                                   &
   2.2$ \, \cdot\, 10^{-8}$                                   &
   2.3$ \, \cdot\, 10^{-6}$                                   &
   1.6$ \, \cdot\, 10^{-3}$                                  \\
     & $10^6$ &
   3.2$ \, \cdot\, 10^{-6}$                                   &
   1.4$ \, \cdot\, 10^{-4}$                                   &
   3.8$ \, \cdot\, 10^{-8}$                                   &
   3.8$ \, \cdot\, 10^{-6}$                                   &
   2.8$ \, \cdot\, 10^{-3}$                                  \\
  & $10^{10}$ &
   5.1$ \, \cdot\, 10^{-6}$                                   &
   2.2$ \, \cdot\, 10^{-4}$                                   &
   6.0$ \, \cdot\, 10^{-8}$                                   &
   5.8$ \, \cdot\, 10^{-6}$                                   &
   4.3$ \, \cdot\, 10^{-3}$                                  \\
  & $10^{16}$ &
   7.0$ \, \cdot\, 10^{-6}$                                   &
   3.0$ \, \cdot\, 10^{-4}$                                   &
   7.9$ \, \cdot\, 10^{-8}$                                   &
   7.3$ \, \cdot\, 10^{-6}$                                   &
   5.6$ \, \cdot\, 10^{-3}$                                  \\
                                                                                                                                                                                                                                                                                                                                                                                                                                                                                                                                                                          \hline \hline \multirow{4}{*}{110}  & $10^4$ &
   1.8$ \, \cdot\, 10^{-6}$                                   &
   8.2$ \, \cdot\, 10^{-5}$                                   &
   2.2$ \, \cdot\, 10^{-8}$                                   &
   2.3$ \, \cdot\, 10^{-6}$                                   &
   1.6$ \, \cdot\, 10^{-3}$                                  \\
     & $10^6$ &
   3.2$ \, \cdot\, 10^{-6}$                                   &
   1.4$ \, \cdot\, 10^{-4}$                                   &
   3.8$ \, \cdot\, 10^{-8}$                                   &
   3.8$ \, \cdot\, 10^{-6}$                                   &
   2.7$ \, \cdot\, 10^{-3}$                                  \\
  & $10^{10}$ &
   5.1$ \, \cdot\, 10^{-6}$                                   &
   2.3$ \, \cdot\, 10^{-4}$                                   &
   6.0$ \, \cdot\, 10^{-8}$                                   &
   5.8$ \, \cdot\, 10^{-6}$                                   &
   4.1$ \, \cdot\, 10^{-3}$                                  \\
  & $10^{16}$ &
   7.1$ \, \cdot\, 10^{-6}$                                   &
   3.0$ \, \cdot\, 10^{-4}$                                   &
   8.0$ \, \cdot\, 10^{-8}$                                   &
   7.4$ \, \cdot\, 10^{-6}$                                   &
   5.4$ \, \cdot\, 10^{-3}$                                  \\
                                                                                                                                                                                                                                                                                                                                                                                                                                                                                                                                                                          \hline \hline \multirow{4}{*}{120}  & $10^4$ &
   1.8$ \, \cdot\, 10^{-6}$                                   &
   8.1$ \, \cdot\, 10^{-5}$                                   &
   2.2$ \, \cdot\, 10^{-8}$                                   &
   2.2$ \, \cdot\, 10^{-6}$                                   &
   1.5$ \, \cdot\, 10^{-3}$                                  \\
     & $10^6$ &
   3.2$ \, \cdot\, 10^{-6}$                                   &
   1.4$ \, \cdot\, 10^{-4}$                                   &
   3.8$ \, \cdot\, 10^{-8}$                                   &
   3.8$ \, \cdot\, 10^{-6}$                                   &
   2.5$ \, \cdot\, 10^{-3}$                                  \\
  & $10^{10}$ &
   5.2$ \, \cdot\, 10^{-6}$                                   &
   2.3$ \, \cdot\, 10^{-4}$                                   &
   6.0$ \, \cdot\, 10^{-8}$                                   &
   5.8$ \, \cdot\, 10^{-6}$                                   &
   4.0$ \, \cdot\, 10^{-3}$                                  \\
  & $10^{16}$ &
   7.2$ \, \cdot\, 10^{-6}$                                   &
   3.1$ \, \cdot\, 10^{-4}$                                   &
   8.1$ \, \cdot\, 10^{-8}$                                   &
   7.5$ \, \cdot\, 10^{-6}$                                   &
   5.3$ \, \cdot\, 10^{-3}$                                  \\
                                                                                                                                                                                                                                                                                                                                                                                                                                                                                                                                                                          \hline \hline \multirow{4}{*}{130}  & $10^4$ &
   1.7$ \, \cdot\, 10^{-6}$                                   &
   8.0$ \, \cdot\, 10^{-5}$                                   &
   2.2$ \, \cdot\, 10^{-8}$                                   &
   2.2$ \, \cdot\, 10^{-6}$                                   &
   1.4$ \, \cdot\, 10^{-3}$                                  \\
     & $10^6$ &
   3.2$ \, \cdot\, 10^{-6}$                                   &
   1.4$ \, \cdot\, 10^{-4}$                                   &
   3.8$ \, \cdot\, 10^{-8}$                                   &
   3.8$ \, \cdot\, 10^{-6}$                                   &
   2.4$ \, \cdot\, 10^{-3}$                                  \\
  & $10^{10}$ &
   5.2$ \, \cdot\, 10^{-6}$                                   &
   2.3$ \, \cdot\, 10^{-4}$                                   &
   6.1$ \, \cdot\, 10^{-8}$                                   &
   5.9$ \, \cdot\, 10^{-6}$                                   &
   3.8$ \, \cdot\, 10^{-3}$                                  \\
  & $10^{16}$ &
   7.3$ \, \cdot\, 10^{-6}$                                   &
   3.1$ \, \cdot\, 10^{-4}$                                   &
   8.2$ \, \cdot\, 10^{-8}$                                   &
   7.6$ \, \cdot\, 10^{-6}$                                   &
   5.1$ \, \cdot\, 10^{-3}$                                  \\
                                                                                                                                                                                                                                                                                                                                                                                                                                                                                                                                                                          \hline \hline \multirow{4}{*}{140}  & $10^4$ &
   1.7$ \, \cdot\, 10^{-6}$                                   &
   7.9$ \, \cdot\, 10^{-5}$                                   &
   2.1$ \, \cdot\, 10^{-8}$                                   &
   2.2$ \, \cdot\, 10^{-6}$                                   &
   1.3$ \, \cdot\, 10^{-3}$                                  \\
     & $10^6$ &
   3.2$ \, \cdot\, 10^{-6}$                                   &
   1.4$ \, \cdot\, 10^{-4}$                                   &
   3.8$ \, \cdot\, 10^{-8}$                                   &
   3.8$ \, \cdot\, 10^{-6}$                                   &
   2.3$ \, \cdot\, 10^{-3}$                                  \\
  & $10^{10}$ &
   5.3$ \, \cdot\, 10^{-6}$                                   &
   2.3$ \, \cdot\, 10^{-4}$                                   &
   6.1$ \, \cdot\, 10^{-8}$                                   &
   5.9$ \, \cdot\, 10^{-6}$                                   &
   3.6$ \, \cdot\, 10^{-3}$                                  \\
  & $10^{16}$ &
   7.4$ \, \cdot\, 10^{-6}$                                   &
   3.2$ \, \cdot\, 10^{-4}$                                   &
   8.4$ \, \cdot\, 10^{-8}$                                   &
   7.7$ \, \cdot\, 10^{-6}$                                   &
   4.9$ \, \cdot\, 10^{-3}$                                  \\
                                                                                                                                                                                                                                                                                                                                                                                                                                                                                                                                                                          \hline \hline \multirow{4}{*}{150}  & $10^4$ &
   1.7$ \, \cdot\, 10^{-6}$                                   &
   7.8$ \, \cdot\, 10^{-5}$                                   &
   2.1$ \, \cdot\, 10^{-8}$                                   &
   2.2$ \, \cdot\, 10^{-6}$                                   &
   1.2$ \, \cdot\, 10^{-3}$                                  \\
     & $10^6$ &
   3.2$ \, \cdot\, 10^{-6}$                                   &
   1.4$ \, \cdot\, 10^{-4}$                                   &
   3.8$ \, \cdot\, 10^{-8}$                                   &
   3.8$ \, \cdot\, 10^{-6}$                                   &
   2.1$ \, \cdot\, 10^{-3}$                                  \\
  & $10^{10}$ &
   5.3$ \, \cdot\, 10^{-6}$                                   &
   2.3$ \, \cdot\, 10^{-4}$                                   &
   6.2$ \, \cdot\, 10^{-8}$                                   &
   6.0$ \, \cdot\, 10^{-6}$                                   &
   3.4$ \, \cdot\, 10^{-3}$                                  \\
  & $10^{16}$ &
   7.6$ \, \cdot\, 10^{-6}$                                   &
   3.2$ \, \cdot\, 10^{-4}$                                   &
   8.5$ \, \cdot\, 10^{-8}$                                   &
   7.9$ \, \cdot\, 10^{-6}$                                   &
   4.6$ \, \cdot\, 10^{-3}$                                  \\
                                                                                                                                                                                                                                                                                                                                                                                                                                                                                                                                                                          \hline \hline \multirow{4}{*}{160}  & $10^4$ &
   1.7$ \, \cdot\, 10^{-6}$                                   &
   7.7$ \, \cdot\, 10^{-5}$                                   &
   2.1$ \, \cdot\, 10^{-8}$                                   &
   2.1$ \, \cdot\, 10^{-6}$                                   &
   1.1$ \, \cdot\, 10^{-3}$                                  \\
     & $10^6$ &
   3.2$ \, \cdot\, 10^{-6}$                                   &
   1.4$ \, \cdot\, 10^{-4}$                                   &
   3.8$ \, \cdot\, 10^{-8}$                                   &
   3.8$ \, \cdot\, 10^{-6}$                                   &
   1.9$ \, \cdot\, 10^{-3}$                                  \\
  & $10^{10}$ &
   5.4$ \, \cdot\, 10^{-6}$                                   &
   2.4$ \, \cdot\, 10^{-4}$                                   &
   6.3$ \, \cdot\, 10^{-8}$                                   &
   6.1$ \, \cdot\, 10^{-6}$                                   &
   3.2$ \, \cdot\, 10^{-3}$                                  \\
  & $10^{16}$ &
   7.7$ \, \cdot\, 10^{-6}$                                   &
   3.3$ \, \cdot\, 10^{-4}$                                   &
   8.7$ \, \cdot\, 10^{-8}$                                   &
   8.0$ \, \cdot\, 10^{-6}$                                   &
   4.4$ \, \cdot\, 10^{-3}$             \\ \hline \end{tabular}


\end{center} 
\caption[]{Absolute values of the
effective FC Yukawa couplings $[{\bf Y^{L,R}_D}]_{23}$ and
$[{\bf Y^{L,R}_U}]_{23}$ corresponding to the FC transitions 
$s\leftrightarrow b $ and $c \leftrightarrow	 t$, respectively,
all evaluated at the 
scale $\mu=m_H$.  
The $b$-quark Yukawa coupling   ${\rm Y}_b\equiv {\rm Y_{D_3}}$ is reported for reference in the last column. 
}
\label{tableYFC} 
\end{table}
In Table~\ref{tableYFC}, 
we present the numerical (absolute) values of 
$[{\bf Y^{L,R}_{U,D}}]_{23}$, that are the most significant 
FC Yukawa couplings, 
evaluated  $\mu=m_H$. Because of the equivalence 
$({\bf Y^{L}_{U,D}})^{\dag}={\bf Y^{R}_{U,D}}$,  one has 
$[{\bf Y^{L,R}_{U,D}}]_{32}=[{\bf Y^{R,L}_{U,D}}]_{23}$. Regarding the CKM matrix
elements, in the Wolfenstein parameterization we set $\lambda=0.2253$, $A=0.808$
\cite{PDG2010}.
In the last column of  Table \ref{tableYFC}, we report for 
comparison the effective bottom-quark Yukawa coupling
${\rm Y}_b\equiv {\rm Y_{D_3}}$.
One can see that the coupling $[{\bf Y^R_D}]_{23}$ responsible for the $b\leftrightarrow  s$ transitions is the largest  FC coupling.
This is because  
the leading  contribution to the $\beta$ function is provided 
by the 2-$W$ exchange diagram in Fig.6(a). Then, 
the GIM mechanism 
makes the $b\leftrightarrow s$ 
transition amplitude ${\cal O}(m_t^2/M_W^2)$, corresponding to a top-quark exchange in the loop, while the $t\leftrightarrow c $ 
transition is depleted by  ${\cal O}(m_b^2/M_W^2)$.
Note that, in  all the range of parameters 100 GeV$\lsim m_H\lsim 160$ GeV and  $\Lambda\sim 10^{(4-16)}$ GeV, one has 
$[{\bf Y^R_D}]_{23}\gsim{\rm Y}_b/20$.

In Table~\ref{tableYFC}, one can also check that  the right-handed couplings mediating the transition between the second  and third  family are both dominant, that is 
$[{\bf Y^R_{U,D}}]_{23} > [{\bf Y^L_{U,D}}]_{23}$.
The divergent part of diagram in Fig.6(a) 
is always proportional to the external quark (pole) masses, since, because of chirality suppression, it needs an external fermion mass insertion.
Then, the V-A structure of weak interactions makes the 
$\beta$ functions of $[{\bf Y^R_D}]_{23}$  and $[{\bf Y^L_D}]_{23}$ 
 proportional to the $b$-quark and $s$-quark mass, respectively, 
 and  the 
$\beta$ functions of $[{\bf Y^R_U}]_{23}$  and $[{\bf Y^L_U}]_{23}$ 
 proportional to the $t$-quark and $c$-quark mass, respectively,
which explains the observed hierarchy.

\section{Flavor-changing decay branching ratios}

Before studying the branching ratios for FC Higgs-boson decays $H\to f_if_j$,  we briefly discuss the constraints on the FC Yukawa couplings imposed by flavor-changing neutral-current (FCNC) processes.
FC Higgs-boson interactions 
can induce effective FCNC interactions mediated by local four-fermion
operators, through tree-level Higgs boson exchange \cite{reina}.
Were these interactions  strong enough, they would spoil the  agreement between the 
SM predictions and experimental measurements for the mass splitting 
$\Delta M_q \equiv M_{B^H_q}-M_{B^L_q}$,  where $M_{B^H_q}$
($M_{B^L_q}$) is the heavy (light) mass eigenstate of the $B^0_q-\bar{B}_q^0$ 
meson system, with $q=s,d$.
Starting from the Lagrangian in Eq.(\ref{LFCY}), the contribution of the 
tree-level Higgs-mediated FCNC to the mass splitting  
$\Delta M_s$ is given by
\cite{reina}
\bea
\Delta M_s &=& \frac{5|[{\bf Y_D^R}]_{23}|^2\, f^2_{B} M^3_{B^0_s}}{
24\, m_H^2 (m_b+m_s)^2}\, ,
\label{bound}
\eea
where $f_{B_s}$ and $M_{B^0_s}$ are the decay constant and 
mass of the $B_s^0$ meson
state, and $m_b$ and $m_s$ are pole quark masses.
In Eq.(\ref{bound}), we kept only the leading $|[{\bf Y_D^R}]_{23}|^2$ term, and 
estimated the hadronic matrix element
\bea
\langle B^0_s | [\bar{b}(1-\gamma_5) s] [\bar{b}(1-\gamma_5) s] 
|\bar{B}^0_s \rangle&=&-\frac{5 f_{B_s}^2 M_{B_s^0}^4\, 
{\rm B}_{B_s}}{3(m_b+m_s)^2}
\eea
in the vacuum insertion approximation, with  
${\rm B}_{B_s}=1$ \cite{reina}.
Then, if we require that the Higgs-mediated contribution to 
$\Delta M_{s}$ does not exceed its experimental 
central value $\Delta M^{\rm exp}_{s}=117.0 \times10^{-13}$ GeV \cite{PDG2010}, we get 
\bea
|[{\bf Y^R_D}]_{23}| \lsim 1.5 \times 10^{-3} \left(\frac{m_H[{\rm GeV}]}{120}\right)\, 
\label{boundexp}
\eea
where, for the $B^0_s$ decay constant and mass, we assume 
$f_{B_s}=238.8$ MeV \cite{laiho}, 
$M_{B^0_s}=5.366$ GeV, respectively  \cite{PDG2010}, while
other SM inputs are given in \cite{our}.

We can see that
 $|[{\bf Y^R_D}]_{23}|$  values in Table \ref{tableYFC} are 
well below  the upper bound in Eq.(\ref{boundexp}).
We conclude that the  experimental constraints on $\Delta M_{s}$ do not pose 
any restriction on the allowed $\Lambda$ range\footnote{
Note that the measured value of $\Delta M_{s}$ is in 
good agreement with the SM predictions, that are anyhow affected by large theoretical uncertainties. If one requires that the 
new-physics (NP) contribution to  $\Delta M_s$ does not exceed the difference between the SM prediction and the measured value within $1 \sigma$, one obtains 
$|\Delta M_s^{\rm(NP)}| < 17.3\times 10^{-13}$ GeV 
\cite{golowich}, that would imply $|[{\bf Y^R_D}]_{23}|< 5.8 \times 10^{-4}(\frac{m_H[{\rm GeV}]}{120})$. 
Although less conservative than  Eq.(\ref{boundexp}), this bound   is still consistent with  all values of $|[{\bf Y^R_D}]_{23}|$
in Table \ref{tableYFC}.}.
The same  holds for the constraints on $\Delta M_{d}$, coming from  the 
neutral $B_d^0-\bar{B_d^0}$ system.

We now compute the Higgs-boson width corresponding to the inclusive  decay
$H\to b s$.  
Neglecting the $s$-quark mass effects, we have 
\bea
\Gamma(H\to bs )&=&
\frac{N_c m_H \left(|[{\bf Y^L_D}]_{23}|^2+|[{\bf Y^R_D}]_{23}|^2\right)
}{16 \pi}\left(1-\frac{m_b^2}{m_H^2}
\right)^{3/2}\; ,
\label{GammaHbs}
\eea
where $m_b$ is the $b$-quark pole mass, the FC Yukawa couplings  are  evaluated at the scale $m_H$, and 
$\Gamma(H\to bs )\equiv \Gamma(H\to \bar{b}s )+\Gamma(H\to \bar{s}b )$.

Correspondingly, in Table \ref{tableBR} we show the numerical results for the branching ratio 
${\rm BR}(H\to bs)$ for different $m_H$ and  $\Lambda$ values.
\begin{table}[htbp]
\begin{center}

                                                                                                                                                                                                                                                      \begin{tabular}{|c||c|c|c|c|c|c|}                                   \hline ${\rm \Lambda({\rm GeV})}$                               & ${\rm BR}_{H\to bs}^{100}$ &                                    ${\rm BR}_{H\to bs}^{110}$  &                                     ${\rm BR}_{H\to bs}^{120}$ &                                      ${\rm BR}_{H\to bs}^{130}$ &                                      ${\rm BR}_{H\to bs}^{140}$ &                                      ${\rm BR}_{H\to bs}^{150}$                                        \\ \hline \hline $10^4$ &     7.7$ \, \cdot\, 10^{-4}$          &    2.5$ \, \cdot\, 10^{-4}$          &    8.1$ \, \cdot\, 10^{-5}$          &    2.9$ \, \cdot\, 10^{-5}$          &    1.1$ \, \cdot\, 10^{-5}$          &    4.1$ \, \cdot\, 10^{-6}$            \\ \hline
                                                                                                                                                                                                                                                                                                                                                                                                                                                                                                           $10^6$ &     1.5$ \, \cdot\, 10^{-3}$          &    6.5$ \, \cdot\, 10^{-4}$          &    2.4$ \, \cdot\, 10^{-4}$          &    9.0$ \, \cdot\, 10^{-5}$          &    3.6$ \, \cdot\, 10^{-5}$          &    1.4$ \, \cdot\, 10^{-5}$                                                                                                                \\ \hline
                                                                                                                                                                                                                                                                                                                                                                                                                                                                                                        $10^{10}$ &     2.1$ \, \cdot\, 10^{-3}$          &    1.2$ \, \cdot\, 10^{-3}$          &    5.5$ \, \cdot\, 10^{-4}$          &    2.3$ \, \cdot\, 10^{-4}$          &    9.5$ \, \cdot\, 10^{-5}$          &    3.7$ \, \cdot\, 10^{-5}$                                                                                                                \\ \hline
                                                                                                                                                                                                                                                                                                                                                                                                                                                                                                        $10^{16}$ &     2.4$ \, \cdot\, 10^{-3}$          &    1.7$ \, \cdot\, 10^{-3}$          &    8.9$ \, \cdot\, 10^{-4}$          &    4.0$ \, \cdot\, 10^{-4}$          &    1.8$ \, \cdot\, 10^{-4}$          &    7.0$ \, \cdot\, 10^{-5}$                                                                                                  \\ \hline \end{tabular}

\end{center} 
\caption[]{Branching ratio ${\rm BR}^{m_H}$ for  $H\to bs$, 
versus $m_H$ (in GeV), and the  energy scale $\Lambda$.}
\label{tableBR} 
\end{table}
\begin{table}[htbp]
\begin{center}
                                                                                                                                                                                                                                                       \begin{tabular}{|c||c|c|c|c|c|c|}                                   \hline ${\rm \Lambda({\rm GeV})}$                               & $N_{\rm ev}^{100}(bs)$ &                                        $N_{\rm ev}^{110}(bs)$  &                                         $N_{\rm ev}^{120}(bs)$ &                                          $N_{\rm ev}^{130}(bs)$ &                                          $N_{\rm ev}^{140}(bs)$ &                                          $N_{\rm ev}^{150}(bs)$                                            \\ \hline \hline $10^4$ &     57          &    18         &    5.4          &    1.8          &    0.6          &    0.2 \\ \hline
                                                                                                                                                                                                                                                                                                                                                                                                                                                                                                           $10^6$ &     110  &    45         &    16           &    5.6          &    2.1          &    0.7                                                                                                             \\ \hline
                                                                                                                                                                                                                                                                                                                                                                                                                                                                                                        $10^{10}$ &     150           &    86         &    36      &    14          &    5.6          &    2.0                                                                                                                \\ \hline
                                                                                                                                                                                                                                                                                                                                                                                                                                                                                                        $10^{16}$ &     180          &    120     &    59   &    25    &     10           &    3.8                                                                                                  \\ \hline \end{tabular}

\end{center} 
\caption[]{Number of expected events $N_{\rm ev}^{m_H}(bs)$
for an integrated luminosity of 500 fb$^{-1}$, corresponding to 
$e^+e^-\to Z H\to Z bs$ at $\sqrt{S}\simeq 350$ GeV, 
versus $m_H$ (in GeV), and the scale $\Lambda$.}
\label{tableNEVFC} 
\end{table}
We can see that the ${\rm BR}(H\to bs)$ can be as large as 
 ${\cal O}(10^{-3})$  for  $m_H\lsim 110$ GeV,
and $\Lambda$ large enough.
Values up to ${\cal O}(10^{-4})$ can be obtained also for 
$m_H\lsim 140$ GeV. In Fig.\ref{fig2},
 ${\rm BR}(H\to bs)$ versus the scale $\Lambda$ is 
plotted, 
for $10^4~ {\rm GeV} < \Lambda < 10^{16}~ {\rm GeV}$, and
for $m_H=120$ GeV (left) and 140 GeV (right).

Note that ${\rm BR}(H\to bs)$ turns out to be almost comparable to
BR$(H\to c\bar c)$ and BR$(H\to\tau\tau)$ for 
$m_H\lsim 120$ GeV (cf. Table \ref{table1}). A measurement of  BR$(H\to bs)$ would  then be feasible at a linear collider. This is  in contrast with what can be achieved at the LHC, where hadronic final states produced through EW processes are typically very challenging, even for unsuppressed couplings.

\begin{figure}[tpb]
\begin{center}
\dofigB{3.1in}{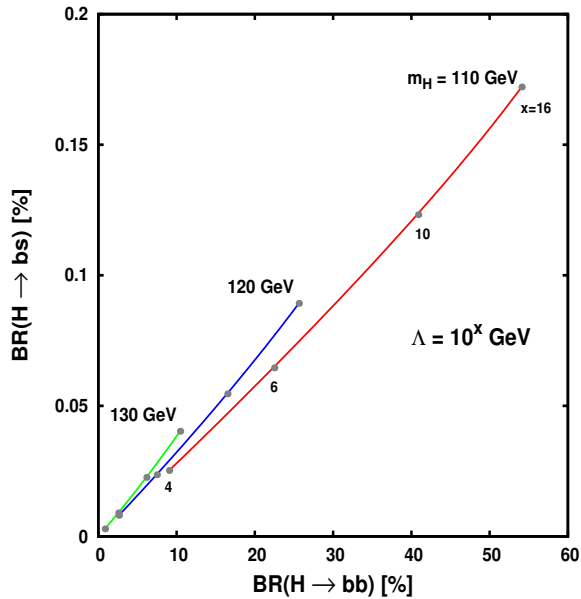}
\end{center}
\caption{\small 
Correlations  between BR$(H\to bs)$ and BR$(H\to b\bar b)$ for different  $m_H$ values.
 }
\label{fig7}
\end{figure}
 In Fig.\ref{fig7}, we show  correlations between BR$(H\to bs)$
and BR$(H\to b\bar b)$,  for 
 $m_H=$110, 120, 130 GeV. 
 For each $m_H$ value, we show by grey bubbles the points corresponding to $\Lambda= 10^{4}, 10^{6},10^{10},10^{16}$ GeV, that are  
 univocally set by  
BR$(H\to b\bar b)$, for any given $m_H$. 
Note that the $m_H$ dependence in the slopes is much reduced with respect to the flavor-diagonal decay correlations in 
Fig.\ref{fig5} (right). This is because, in the  Eqs.(\ref{betaUL})-(\ref{betaDR}) for the FC couplings, the dependence on 
$\xi_H \propto m_H^2/m_W^2$ is subdominant (i.e., depleted by radiative couplings) with respect to the Eqs.(\ref{RGE1})-(\ref{RGE3}) for the flavor-diagonal couplings, 
 where $\xi_H$ terms are enhanced by the ChSB fermion masses.

In Table \ref{tableNEVFC}, we report the expected number of events $N_{\rm ev}^{m_H}(bs)$ for the 
FC Higgs decay $H\to bs$,
corresponding to the production channel 
$e^+e^-\to Z H\to Z bs$ at $\sqrt{S}\simeq 350$ GeV and
with integrated luminosity of 500 fb$^{-1}$.
One can see that 18 (120)
$Zbs$ events are expected for  $m_H=110$ and
$\Lambda=10^{4\, (16)}$GeV,  decreasing to 2.1 (10)
for $m_H=140$ and $\Lambda=10^{6\, (16)}$GeV.
Considering the moderate background environment of the linear collider, we then expect that a 
 detailed study 
 including backgrounds   and detection efficiencies could confirm the possibility of making a measurement of BR$(H\to b s)$ for a quite wide range of the model parameters.

\section{Conclusions}

In this paper, we examined the potential of a linear collider program
for testing the effective Yukawa scenario. With respect to the SM, this theoretical framework is characterized by a Higgs boson with  radiative (and hence depleted) Yukawa couplings to fermions, and
unaltered couplings to EW massive vector bosons.  
 LHC will be able to pinpoint this scenario that,  at hadron colliders,   for  $m_H\lsim 150$ GeV, foresees a  Higgs boson mainly produced by vector-boson fusion with SM cross sections, with enhanced decays to $\gamma\gamma,WW,ZZ$. 
The direct investigation of the fermionic sector of the Higgs-boson couplings requires instead the clean environment of a linear collider program. We showed, that, with a typical $e^+e^-$ setup
with  $\sqrt{S}\simeq 350$ GeV and 500 fb$^{-1}$ of integrated luminosity, the production rates for a Higgs boson decaying 
into $b\bar b, c\bar c, \tau\tau$ are sufficiently large to allow
a nice determination of the corresponding effective Yukawa couplings, for 
 $m_H\lsim 150$ GeV. Furthermore, since fermionic BR's are particularly sensitive to the large energy scale $\Lambda$ (where the Yukawa couplings are assumed to vanish at tree level), a measurement  of the high-statistic channel  $H\to b\bar b$ is expected to provide  a good $\Lambda$ determination  even for $m_H\gsim 120$ GeV, where the sensitivity to the scale $\Lambda$ of BR$(H\to \gamma\gamma,WW,ZZ)$ decreases. Another sector where LHC can not compete with a linear collider is the study of the enhanced FC Higgs-boson  decay $H\to bs$, for which the low hadronic background of a linear collider is vital for detection. 
 In particular, we showed that BR$(H\to bs)\sim(10^{-4}-10^{-3})$, that is almost of the same order of 
 BR$(H\to c\bar c)$ and BR$(H\to  \tau\tau)$, is expected for
 $m_H\lsim 120$, with a corresponding event statistic sufficient
 for a nice  BR$(H\to bs)$ determination.
 More detailed conclusions will require a more refined  phenomenological analysis including the relevant backgrounds   and detection efficiencies.

\section*{\bf Acknowledgments}
We would like to thank Gad Eilam for pointing out reference
\cite{Arhrib}.
B.M. was partially supported by the RTN European 
Programme Contract No. MRTN-CT-2006-035505 (HEPTOOLS, Tools and Precision Calculations 
for Physics Discoveries at Colliders).


\newpage
\begin{thebibliography}{99}
\bibitem{Djouadi}
For a review see e.g. 
  A.~Djouadi,
  Phys.\ Rept.\  {\bf 457}, 1 (2008),
  arXiv:hep-ph/0503172.
\bibitem{LC}
J.~A.~Aguilar-Saavedra {\it et al.}  [ECFA/DESY LC Physics Working Group],
``TESLA Technical Design Report Part III: Physics at an e+e- Linear Collider'', 
R.D.Heuer, D.Miller, F.Richard, P.Zerwas, Eds., 
  arXiv:hep-ph/0106315, 
   http://tesla.desy.de/new$\_$pages/TDR$\_$CD/start.html  .
\bibitem{clic}
  E.~Accomando {\it et al.}  [CLIC Physics Working Group],
  ``Physics at the CLIC multi-TeV linear collider'', 
  M. Battaglia, A. De Roeck, J. Ellis, D. Schulte, Eds., 
  CERN-2004-005, 
  arXiv:hep-ph/0412251.
\bibitem{our}
  E.~Gabrielli and B.~Mele,
  Phys.\ Rev.\  D {\bf 82}, 113014 (2010), 
  arXiv:1005.2498 [hep-ph].
  \bibitem{HiggsFP}
  See, for instance, 
  H.~E.~Haber, G.~L.~Kane and T.~Sterling,
  Nucl.\ Phys.\  B {\bf 161},  493 (1979);
  J.~F.~Gunion, R.~Vega and J.~Wudka,
  Phys.\ Rev.\  D {\bf 42},  1673 (1990);
  P.~Bamert and Z.~Kunszt,
  Phys.\ Lett.\  B {\bf 306},  335 (1993), 
  arXiv:hep-ph/9303239;
  A.~G.~Akeroyd,
  Phys.\ Lett.\  B {\bf 368}, 89 (1996),
  arXiv:hep-ph/9511347;
  A.~Barroso, L.~Brucher and R.~Santos,
  Phys.\ Rev.\  D {\bf 60},  035005 (1999), 
  arXiv:hep-ph/9901293; 
  L.~Brucher and R.~Santos,
  Eur.\ Phys.\ J.\  C {\bf 12}, 87 (2000), 
  arXiv:hep-ph/9907434.
\bibitem{LEP}
R. Barate et al. (LEP Working Group for Higgs Boson 
Searches and ALEPH Collaboration), Phys. Lett. B 565, 
61 (2003), arXiv:hep-ex/0306033.  
\bibitem{Tevatron}
CDF and D0 collaborations, Phys. Rev. Lett. 104,
 061802 (2010); updated in arXiv:1007.4587 [hep-ex].
\bibitem{djou}
G.~Aarons {\it et al.} 
[ILC Collaboration],
  ``International Linear Collider Reference Design Report, Volume 2: PHYSICS AT  THE ILC'',
  A.Djouadi, J.Lykken, K.M\"onig, Y.Okada, M.Oreglia, S.Yamashita, Eds.,
  arXiv:0709.1893 [hep-ph].
\bibitem{batt}
  M.~Battaglia, 
  ``The International Linear Collider'',
  arXiv:0705.3997 [hep-ex].
\bibitem{Battaglia:1999re}
  M.~Battaglia,
  ``Measuring Higgs branching ratios and telling the SM from a MSSM Higgs boson
  at the e+ e- linear collider'', 
  arXiv:hep-ph/9910271.
 \bibitem{Eilam}
 G.~Eilam, B.~Haeri and A.~Soni,
  Phys.\ Rev.\  D {\bf 41}, 875 (1990).
 \bibitem{Arhrib}
 A.~Arhrib,
  Phys.\ Lett.\  B {\bf 612}, 263 (2005), 
  arXiv:hep-ph/0409218.
\bibitem{HbsSUSY}
  A.~M.~Curiel, M.~J.~Herrero and D.~Temes,
  Phys.\ Rev.\  D {\bf 67}, 075008 (2003), 
  arXiv:hep-ph/0210335;
  A.~M.~Curiel, M.~J.~Herrero, W.~Hollik, F.~Merz and S.~Penaranda,
  Phys.\ Rev.\  D {\bf 69}, 075009 (2004), 
  arXiv:hep-ph/0312135;
  S.~Bejar, J.~Guasch and J.~Sola,
  JHEP {\bf 0510}, 113 (2005), 
 arXiv:hep-ph/0508043;
  W.~Hollik, S.~Penaranda and M.~Vogt,
  Eur.\ Phys.\ J.\  C {\bf 47}, 207 (2006),
  arXiv:hep-ph/0511021;
  S.~Bejar, F.~Dilme, J.~Guasch and J.~Sola,
  JHEP {\bf 0408}, 018 (2004),
  arXiv:hep-ph/0402188; 
  A.~Arhrib, D.~K.~Ghosh, O.~C.~W.~Kong and R.~D.~Vaidya,
  Phys.\ Lett.\  B {\bf 647}, 36 (2007), 
  arXiv:hep-ph/0605056;
  S.~Bejar, J.~Guasch, D.~Lopez-Val and J.~Sola,
  Phys.\ Lett.\  B {\bf 668}, 364 (2008), 
 arXiv:0805.0973 [hep-ph].
\bibitem{arason}
  H.~Arason, D.~J.~Castano, B.~Keszthelyi, S.~Mikaelian, E.~J.~Piard, P.~Ramond and B.~D.~Wright,
  Phys.\ Rev.\  D {\bf 46} (1992) 3945.
\bibitem{reina}
  D.~Atwood, L.~Reina and A.~Soni,
  Phys.\ Rev.\  D {\bf 55}, 3156 (1997), 
  arXiv:hep-ph/9609279.
\bibitem{PDG2010}
  K.~Nakamura {\it et al.}  [Particle Data Group],
  ``Review of particle physics'',
  J.\ Phys.\ G {\bf 37}, 075021 (2010).
\bibitem{laiho}
  J.~Laiho, E.~Lunghi and R.~S.~Van de Water,
  Phys.\ Rev.\  D {\bf 81}, 034503 (2010),
  arXiv:0910.2928 [hep-ph].
\bibitem{golowich}
  E.~Golowich, J.~Hewett, S.~Pakvasa, A.~A.~Petrov and G.~K.~Yeghiyan,
  arXiv:1102.0009 [hep-ph].

\end{thebibliography}
\end{document}